\documentclass[two column]{aastex63}
\usepackage{natbib}
\usepackage{booktabs}
\usepackage{graphicx}
\usepackage{multirow}
\usepackage{xcolor}

\received{June 21, 2021}
\revised{August 29, 2021}
\accepted{September 14, 2021}

\submitjournal{ApJ}

\shorttitle{2-kpc Sized Environments of Supernovae}
\shortauthors{Cronin, Utomo et al.}

\begin{document}

\title{Local Environments of Low-Redshift Supernovae} 

\author[0000-0002-9511-1330]{Serena A. Cronin}
\affiliation{Department of Astronomy, The Ohio State University, 4055 McPherson Laboratory, 140 West 18th Avenue, Columbus, OH 43210, USA}
\affiliation{Department of Astronomy, University of Maryland, College Park, MD 20742, USA}
\affiliation{National Radio Astronomy Observatory,
1003 Lopezville Road, Socorro, NM 87801, USA}
\author[0000-0003-4161-2639]{Dyas Utomo}\thanks{Jansky Fellow of the National Radio Astronomy Observatory}
\affiliation{National Radio Astronomy Observatory, 520 Edgemont Rd., Charlottesville, VA 22903, USA}
\affiliation{Department of Astronomy, The Ohio State University, 4055 McPherson Laboratory, 140 West 18th Avenue, Columbus, OH 43210, USA}
\affiliation{Center for Cosmology and AstroParticle Physics, 191 West Woodruff Avenue, Columbus, OH 43210, USA}
\author[0000-0002-2545-1700]{Adam K. Leroy}
\affiliation{Department of Astronomy, The Ohio State University, 4055 McPherson Laboratory, 140 West 18th Avenue, Columbus, OH 43210, USA}
\affiliation{Center for Cosmology and AstroParticle Physics, 191 West Woodruff Avenue, Columbus, OH 43210, USA}
\author[0000-0002-2333-5474]{Erica A. Behrens}
\affiliation{Department of Astronomy, University of Virginia, 530 McCormick Rd., Charlottesville, VA, 22904, USA}
\affiliation{Department of Astronomy, The Ohio State University, 4055 McPherson Laboratory, 140 West 18th Avenue, Columbus, OH 43210, USA}
\author[0000-0002-5235-5589]{J\'{e}r\'{e}my Chastenet}
\affiliation{Sterrenkundig Observatorium, Ghent University, Kri-jgslaan 281 $-$ S9, 9000 Gent, Belgium}
\author[0000-0002-7643-0504]{Tyler Holland-Ashford}
\affiliation{Department of Astronomy, The Ohio State University, 4055 McPherson Laboratory, 140 West 18th Avenue, Columbus, OH 43210, USA}
\affiliation{Center for Cosmology and AstroParticle Physics, 191 West Woodruff Avenue, Columbus, OH 43210, USA}
\author[0000-0001-9605-780X]{Eric W. Koch}
\affiliation{Center for Astrophysics $|$ Harvard \& Smithsonian, 60 Garden St., 02138 Cambridge, MA, USA}
\author[0000-0002-1790-3148]{Laura A. Lopez}
\affiliation{Department of Astronomy, The Ohio State University, 4055 McPherson Laboratory, 140 West 18th Avenue, Columbus, OH 43210, USA}
\affiliation{Center for Cosmology and AstroParticle Physics, 191 West Woodruff Avenue, Columbus, OH 43210, USA}
\author[0000-0002-4378-8534]{Karin M. Sandstrom}
\affiliation{Center for Astrophysics and Space Sciences, Department of Physics, University of California, San Diego
9500 Gilman Drive, La Jolla, CA 92093, USA}
\author[0000-0002-0012-2142]{Thomas G. Williams} \affiliation{Max Planck Institut f{\"u}r Astronomie, K{\"o}nigstuhl 17, 69117 Heidelberg, Germany}

\correspondingauthor{Serena A. Cronin}
\email{cronin@umd.edu}

\begin{abstract}

We characterize the local (2-kpc sized) environments of Type Ia, II, and Ib/c supernovae (SNe) that have recently occurred in nearby ($d\lesssim50$~Mpc) galaxies. Using ultraviolet (UV, from GALEX) and infrared (IR, from WISE) maps of $359$ galaxies and a sample of $472$ SNe, we measure the star formation rate surface density ($\Sigma_{\rm SFR}$) and stellar mass surface density ($\Sigma_\star$) in a 2-kpc beam centered on each SN site. We show that core-collapse SNe are preferentially located along the resolved galactic star-forming main sequence, whereas Type Ia SNe are extended to lower values of $\Sigma_{\rm SFR}$ at fixed $\Sigma_\star$, indicative of locations inside quiescent galaxies or quiescent regions of galaxies. We also test how well the radial distribution of each SN type matches the radial distributions of UV and IR light in each host galaxy. We find that, to first order, the distributions of all types of SNe mirror that of both near-IR light (3.4 and 4.5~\micron, tracing the stellar mass distribution) and mid-IR light (12 and 22~\micron, tracing emission from hot, small grains), and also resemble our best-estimate $\Sigma_{\rm SFR}$. All types of SNe appear more radially concentrated than the UV emission of their host galaxies. In more detail, the distributions of Type II SNe show small statistical differences from that of near-IR light. We attribute this overall structural uniformity to the fact that within any individual galaxy, $\Sigma_{\rm SFR}$ and $\Sigma_\star$ track one another well, with variations in $\Sigma_{\rm SFR}/\Sigma_\star$ most visible when comparing between galaxies.
\end{abstract}


\section{Introduction} \label{sec:intro}

Surveys like the Lick Observatory Supernova Search \citep[LOSS;][]{FILIPPENKO01}, the All-Sky Automatic Survey for Supernovae \citep[ASAS-SN;][]{shappee14,kochanek17}, the Zwicky Transient Facility \citep[ZTF;][]{bellm19}, and the Asteroid Terrestrial-impact Last Alert System \citep[ATLAS;][]{tonry18} have led to a surge in supernova (SN) detections in recent years. The identification of many new supernovae (SNe) has improved our knowledge of their sub-types and locations. Nonetheless, fundamental questions remain about the progenitors of each SN type and the environments in which they explode.

Current theories and observations indicate that Type Ia SNe originate from the thermonuclear explosions of white dwarfs (e.g., see review by \citealt{maoz14}), while core-collapse SNe (Types II, Ib, and Ic) arise when the iron cores of young, massive ($ \gtrsim 8~M_\odot$) stars cannot be supported by nuclear fusion (e.g., \citealt{iben74,muller16}). This picture is supported by the properties of SN host galaxies: all types of SNe II are rare in early-type galaxies, whereas SNe Ia typically occur in all Hubble types \citep{vandenbergh05}. This broad picture leaves room for a much more detailed understanding of, for example, which stars explode, the distribution delay times from star formation to explosion, and how these quantities depend on other factors like local environment or heavy element content. 

Many SNe in nearby galaxies can be localized to a well-characterized environment, where properties like age, mass, recent star formation rate, and local stellar mass density can be inferred from multi-wavelength imaging or even color magnitude diagram fitting of individual stars. Cross-correlating the locations of SNe with these environmental properties allows one to make inferences about SN progenitors and delay time distributions. To carry out this cross-correlation, previous works have adopted a ``pixel statistic'' approach, in which one compares the intensity of host galaxy light at the locations of individual SNe with respect to the overall distribution of light inside the SN host galaxies \citep[{e.g.,~}][]{james06,anderson08,anderson09,kelly08,habergham10,habergham12,galbany12,galbany14,galbany18,anderson12,kangas13,anderson15ia,anderson15,audcent-ross20}. We provide a short summary of some results from these previous works in Section \ref{sec:background}.

So far, most pixel statistic work has focused on optical or ultraviolet emission. In this paper, we extend this work to include infrared (IR) maps of galaxies. To our knowledge, this represents the first systematic measurement of mid-IR emission from the sites of a large sample of SNe. In normal star-forming galaxies, the mid-IR WISE $12~\mu$m and $22~\mu$m emission that we measure arise from small dust grains heated mostly by recent star formation \citep[e.g.,][]{cluver17}. These measurements offer a powerful complement to the tracers of recent star formation used in many previous studies of SN environments, such as H$\alpha$, which can suffer from uncertain extinction effects. In addition to mid-IR emission, we also analyze near-IR WISE $3.4~\mu$m and $4.5~\mu$m, which mainly trace the distribution of stellar mass, and near-UV and far-UV emission, which primarily trace direct emission from massive young stars. We also combine UV and mid-IR emission to estimate a total local star formation rate surface density, $\Sigma_{\rm SFR}$, that includes both obscured and unobscured terms.

To make these measurements, we cross-match the Open Supernova Catalog\footnote{\url{https://sne.space}} \citep[OSC;][]{guillochon17} with an atlas of UV and IR maps of local (distances of $d\lesssim50$~Mpc) galaxies \citep{leroy19}. This allows us to measure the UV and IR emission at the sites of $472$ SNe, comparable in sample size to the largest previous optical study done by \citet[][]{kelly08}. We use these measurements to make two statistical measurements of the local environments of SNe. First, we place the local, 2-kpc resolution environments of SNe into the star formation rate surface density vs. stellar mass surface density plane (i.e.,~$\Sigma_{\rm SFR}{-}\Sigma_{\rm \star}$). We compare the $\Sigma_{\rm SFR}{-}\Sigma_{\rm \star}$ measurements at the locations of SNe with the same measurements at all regions within $r_{25}$ of their host galaxies. That is, we place the environments of different types of SNe relative to the ``resolved star-forming main sequence'' \citep[e.g.,][]{medling18}. Second, we compare the radial distributions of each SN type with the radial distributions of UV, near-IR, and mid-IR light inside their host galaxies. 

This paper is organized as follows. First, we provide background on the results of previous works in Section~\ref{sec:background}. Second, we describe the atlas of UV and IR maps and report how we selected SNe from the OSC in Section~\ref{sec:data}. 
We then characterize the environments of SNe within 2~kpc by placing our SN sample in the $\Sigma_{\rm SFR}{-}\Sigma_{\rm \star}$ plane in Section~\ref{sec:sfms}. To analyze these local environments in more detail, we build radial cumulative distributions that show how each SN type tracks the UV and IR light of their hosts (Section~\ref{sec:method}). We present the resulting SN distributions and analyze the statistical significance of our results in Section~\ref{sec:results}. Lastly, we summarize our results in Section~\ref{sec:conclusions}. Throughout this work, we assume a \citet{chabrier03} initial mass function (IMF).

\subsection{Background}
\label{sec:background}

Many previous pixel statistic studies are reviewed by \citet{anderson15}. We briefly summarize their findings for Type Ia, II, Ib, and Ic SNe to help frame the work in this paper.

{\it SNe~Ia --} \citet{anderson15ia} compared the distribution of Type Ia SNe (SNe Ia) to the distribution of surface brightness in multiple photometric bands within their host galaxies. They found that SNe Ia trace $R$-band light, which generally stems from older stellar populations. Meanwhile, the SNe Ia distribution shows statistically significant differences from H$\alpha$ and NUV emission, which mostly arise from young, massive stars. The better association between SNe Ia and $R-$band is consistent with the idea that SN Ia progenitors are linked to old stellar populations.

{\it SNe~II --} Previous studies suggest that compared to that of SNe Ia, the distribution of Type II SNe (SNe II) offers a better match to tracers of recent star formation and a worse match to tracers of old stellar populations. For example, \citet{anderson12} determined that SNe II follow H$\alpha$ emission better than SNe Ia. They also found that the distribution of SNe II~P, a sub-type of SNe II, matches the distribution of near-UV (NUV) emission, which tracks young, massive stars.

{\it SNe~Ib and Ic --} \citet{anderson12} found that the distribution of Type Ib SNe (SNe Ib) matches that of H$\alpha$ more closely than the distribution of all SNe II matches that of H$\alpha$. Furthermore, Type Ic SNe (SNe Ic) match the distribution of H$\alpha$ even more closely than SNe Ib. This is supported by the findings of \citet{kelly08}, which indicate that SNe Ic are located in the brightest regions of their hosts in {\it g\arcmin}-band. SNe II and Ib also trace {\it g\arcmin}-band light, but are not as concentrated toward the brightest regions as SNe Ic. The authors suggest that these bright regions correspond to the largest star-forming regions, indicating that SNe Ic are more associated with massive star formation than SNe II and Ib.

Using IFU observations from the Calar Alto telescope, \citet{galbany18} studied 272 sites of SNe in 232 host galaxies with an average spatial resolution of 380 pc. They compared the spectra at those SNe sites with respect to the spectra of 11,270 {\sc Hii} regions in host galaxies. They found that SNe Ic occur in {\sc Hii} regions with higher SFR, higher H$\alpha$ equivalent width (EW), and higher metallicity. This indicates that SNe Ic have more massive progenitors than other CC SNe.

Other studies have shown that Type Ib/c SNe (SNe Ib/c; a combination of SNe Ib and SNe Ic) trace recent star formation more closely than SNe II \citep{crowther13,galbany14}. This reflects the understanding that SNe Ib/c progenitors are more massive than SNe II progenitors. As a result, SNe Ib/c tend to occur more often closer to sites of extremely recent star formation, as traced by H$\alpha$.

\section{Data} 
\label{sec:data}

We begin with an atlas of WISE IR and GALEX UV maps of $\sim16,000$ local galaxies \citep{leroy19}. For each galaxy in the atlas, we use the OSC to identify SNe that have occurred in that galaxy. Then, we measure the intensity of IR and UV emission at the location of each SN. We compare these intensities with the distribution of IR and UV intensity from the whole host galaxy. We use these to assess the 2-kpc resolution environments in which SNe Ia, II, and Ib/c explode.

\subsection{Atlas of IR and UV Maps of Galaxies}

\subsubsection{Parent Galaxy Sample} 
\label{sec:galaxies}

We first consider the full set of galaxies with WISE IR and GALEX UV maps from \citet{leroy19}. These maps represent the first part of the $z\approx0$ Multiwavelength Galaxy Synthesis ($z$0MGS), which is an effort to compile a large, multi-wavelength database of galaxies in the local universe. The atlas consists of maps of $\sim$16,000 low-redshift ($z\approx0$) galaxies, $\sim 11,000$ of which are expected to lie within $50$~Mpc and have $M_B < -18$~mag, making them somewhat more luminous than the Large Magellanic Cloud or M33. The atlas is expected to be approximately volume limited for these conditions ($d < 50$~Mpc and $M_B < -18$~mag).

We remove M31 and M33 from the sample due to their large angular size. This removal ensures that we do not falsely place SNe from background galaxies into these fields. This has a minimal impact on our analysis because: 1.)~we do not consider supernova remnants, 2.)~M31 has only one recent SN, and 3.)~M33 has no recent SNe. This is the sole cut we make to the $z$0MGS atlas prior to cross-matching with the OSC.

\subsubsection{Near- and Mid-IR Maps from WISE}
\label{sec:wise}

Our IR maps originate from NASA's WISE telescope \citep{wright10} and make use of a modified version of the unWISE processing \citep{lang14}. The WISE data have four bands, which we refer to as W1, W2, W3, and W4. W1 and W2 are near-IR bands centered at 3.4~$\mu$m and 4.6~$\mu$m. Light in these bands primarily originates from the relatively old stellar population, and W1 in particular is often used to trace stellar mass \citep[e.g., see the summary in][]{leroy19}. Both W1 and W2 can include some contributions from hot dust grains, reflecting the presence of a PAH band in the W1 bandpass and the contribution of hot dust, especially from active galactic nuclei (AGN), in the W2 bandpass \citep[e.g., see][]{wright10,meidt12,querejeta15}.

The W3 and W4 bands are centered near 12~$\mu$m and 22~$\mu$m. Both bands capture mostly emission from hot dust grains. Small dust grains stochastically heated to high temperatures make large contributions to both bands, and the $12~\mu$m band covers several prominent PAH features \citep[e.g., see][]{smith07,wright10}. Ultimately, the emission from these dust grains traces starlight\textemdash mostly UV light, which has been absorbed by dust and then re-emitted. As a result, these bands have a close association with recent star formation and heavily-embedded, massive stars \citep[e.g.,][]{calzetti07,kennicutt12,leroy19}.

\subsubsection{UV Maps From GALEX}
\label{sec:galex}

Our UV maps come from GALEX \citep{martin05} and represent an exposure-time weighted combination of all GALEX images covering each object. GALEX's NUV band is centered near $2{,}300~$\AA,~and the FUV filter is centered near $1{,}500~$\AA. When massive stars are present, they tend to contribute the overwhelming bulk of the light observed in these bands. Dust heavily affects UV emission from galaxies, so the GALEX bands can be thought of as tracing the unobscured light from young, massive stars. Note that while every galaxy in the parent sample has full WISE coverage, only $\sim$2/3 of the targets were covered by GALEX.

\subsubsection{Processing of GALEX and WISE Maps} \label{sec:convolution}
We blank foreground stars and background galaxies in the GALEX and WISE maps using the masks described in \citet{leroy19}. These leverage GAIA DR2 \citep{gaia16,gaia18} and HyperLEDA\footnote{\url{http://leda.univ-lyon1.fr}} \citep{makarov14} to identify the likely footprints of these objects. After blanking contaminated regions, we fill in the affected area with an estimate of the typical galaxy intensity at that galactocentric radius. To do this, we construct a median-based radial profile of intensity in the galaxy, excluding blanked pixels, using bins with width equal to one resolution element. Then, we fill in the blanked pixels with the median value of other pixels at the same radial bin. If all pixels within a radial bin are blank, then we expand the size of the radial bin with a step of one resolution element until the median can be calculated. We carry out this interpolation to fill in all blanked pixels from the center of the galaxy out to $3~r_{25}$, i.e., three times the nominal optical radius of the galaxy.

The $z$0MGS atlas images are at fixed angular resolutions of 15\arcsec~ for W4 and 7.5\arcsec~ for all other bands. Because the targets span more than an order of magnitude in distance, these fixed angular resolutions correspond to a wide range of physical resolutions. In order to make a fair comparison of SN environments between galaxies at different distances, we convolve the maps to a series of fixed physical resolutions. We create versions of the maps that have a full width at half maximum (FWHM) point spread function (PSF) of 0.5~kpc, 1~kpc, and 2~kpc. To do this, we convolve the interpolated maps to an angular resolution such that the FWHM of the PSF would match each value, with the exact target angular resolution set by the distance to the target. This convolution is equivalent to placing all galaxies at the same distance so that each independent resolution element has the same physical size. The equivalent distances to obtaining a PSF of 0.5~kpc, 1~kpc, and 2~kpc from the 15\arcsec\ maps are 6.88~Mpc, 13.75~Mpc, and 27.5~Mpc, respectively.

Because the target angular resolution to reach a fixed physical resolution depends on distance, the number of galaxies that can be convolved to each physical resolution varies. We have far fewer targets at 0.5-kpc and 1-kpc resolution than at 2-kpc resolution. Furthermore, because the W4 maps begin at the coarser resolution of 15\arcsec\ (compared to 7.5\arcsec\ for the other bands), fewer W4 maps can be convolved to any given physical resolution. This results in a noticeably smaller sample size for W4.

\subsubsection{Estimates of the Stellar Mass and Star Formation Rate Surface Densities}
\label{sec:sfr}

We expect CC SNe to track recent star formation. To help test this hypothesis, we calculate two estimations of the SFR per unit area, or SFR surface density: $\Sigma_{\rm SFR}$(FUV$+$W4), which combines measured W4 and GALEX FUV intensities, and $\Sigma_{\rm SFR}$(NUV$+$W3), which combines measured W3 and GALEX NUV intensities. Both WISE and GALEX trace the presence of massive stars, and thus recent star formation, but in complementary ways. While W3 and W4 trace dust-reprocessed UV/optical light, GALEX's NUV and FUV observations capture light from young stars not obscured by dust (as discussed in Sections~\ref{sec:wise} and \ref{sec:galex}). Combining this UV and mid-IR emission allows one to construct a general tracer of star formation \citep[e.g.,][]{meurer99,thilker07,leroy08}.

There are advantages and limitations to using each star formation rate tracer. First, we consider $\Sigma_{\rm SFR}$(FUV$+$W4) to be a more direct tracer of the SFR than $\Sigma_{\rm SFR}$(NUV$+$W3). This is because the W4 band is less likely to be contaminated by PAHs than W3 \citep[PAH abundance varies with metallicity;][]{engelbracht05,engelbracht08,calzetti07} and as a result shows a more stable correlation with the SFR estimated from spectral energy distribution fitting \citep[e.g.,][]{leroy19}. Second, compared to NUV, the FUV band also shows less contamination by light from lower mass and thus potentially older stars. However, the W4 data have poorer resolution and sensitivity than the W3 data, and this poorer resolution implies a smaller sample size when working with W4 at fixed angular resolution. Along similar lines, the NUV sky coverage of GALEX is moderately better than the FUV sky coverage. These considerations limit our ability to estimate $\Sigma_{\rm SFR}$(FUV$+$W4) for the full sample. Therefore, we also include an analysis of $\Sigma_{\rm SFR}$(NUV$+$W3), which allows a much larger sample at the cost of some potential contamination in the SFR tracer.

We follow \citet{leroy19} to convert the WISE and GALEX bands to $\Sigma_{\rm SFR}$. For each pixel in a galaxy, we take the measured intensity and convert it to $\Sigma_{\rm SFR}$ using

\begin{equation}
\label{eqn:w3sfr}
    \frac{\Sigma_{\rm SFR, W3}}{M_{\odot}~{\rm yr^{-1}~kpc^{-2}}} \approx 3.77 \times 10^{-3} \left (  \frac{C_{\rm W3}}{10^{-42.9}}\right ) \left (\frac{I_{\rm 12\mu m }}{{\rm MJy~sr^{-1}}}  \right),
\end{equation}

\begin{equation}
\label{eqn:w4sfr}
    \frac{\Sigma_{\rm SFR,W4}}{M_{\odot}~{\rm yr^{-1}~kpc^{-2}}} \approx 3.24 \times 10^{-3} \left (  \frac{C_{\rm W4}}{10^{-42.7}}\right ) \left (\frac{I_{\rm 22\mu m }}{{\rm MJy~sr^{-1}}}  \right),
\end{equation}

\begin{equation}
\label{eqn:fuvsfr}
    \frac{\Sigma_{\rm SFR,FUV}}{M_{\odot}~{\rm yr^{-1}~kpc^{-2}}} \approx 1.04 \times 10^{-1} \left (\frac{C_{\rm FUV}}{10^{-43.35}}\right ) \left (\frac{I_{\rm FUV}}{{\rm MJy~sr^{-1}}}  \right),
\end{equation}

and

\begin{equation}
\label{eqn:nuvsfr}
    \frac{\Sigma_{\rm SFR, NUV}}{M_{\odot}~{\rm yr^{-1}~kpc^{-2}}} \approx 1.05 \times 10^{-1} \left (  \frac{C_{\rm NUV}}{10^{-43.17}}\right ) \left (\frac{I_{\rm NUV}}{{\rm MJy~sr^{-1}}}  \right),
\end{equation}

\noindent where $C_{\rm W3} = 10^{-42.86}$, $C_{\rm W4} = 10^{-42.73}$, $C_{\rm FUV} = 10^{-43.42}$, and $C_{\rm NUV} = 10^{-43.24}$. We obtain the total $\Sigma_{\rm SFR}$(NUV+W3) as $\Sigma_{\rm SFR,NUV} + \Sigma_{\rm SFR,W3}$ and $\Sigma_{\rm SFR}$(FUV+W4) as $\Sigma_{\rm SFR,FUV} + \Sigma_{\rm SFR,W4}$.

All of the $C$ coefficients in this estimate have been derived to match integrated-galaxy population fitting results for the SDSS main galaxy sample by \citet{salim16}. In addition to statistical noise, we expect $\approx 0.1$~dex systematic uncertainty translating from intensity to $\Sigma_{\rm SFR}$ \citep{leroy19}.

To derive the stellar mass surface density, $\Sigma_{\star}$, from W1, we apply the following conversion, where $M/L$ is the mass-to-light ratio of a galaxy from \citet{leroy19}:

\begin{equation}
\label{eqn:mtol}
    \frac{\Sigma_{\star}}{M_{\odot}~{\rm pc^{-2}}} \approx 3.3 \times 10^{2} \left (  \frac{M/L}{0.5~M_\odot/L_\odot}\right ) \left (\frac{I_{\rm 3.4\mu m }}{{\rm MJy~sr^{-1}}}  \right)~.
\end{equation}

\noindent \citet{leroy19} estimate the mass-to-light ratio based on a prescription using the specific star formation rate. We refer to that paper for details and simply note that here we adopt the single value estimated for each galaxy by \citet{leroy19}, and that older, less active galaxies tend to have higher W1 mass-to-light ratios compared to systems with more active star formation.

\subsection{The Supernova Sample} \label{sec:supernovae}

We use the OSC to identify where recent SNe have occurred in $z$0MGS galaxies. The OSC is a large, heterogeneous collection of aggregated data on all known SNe and SN remnants. Note that the OSC is not a complete SN catalog, because many SNe even in local galaxies will not have been discovered. For example, many early SN surveys were not sensitive to the nuclear regions of galaxies, thus missing SNe that occurred close to the central regions of their hosts \citep[e.g., see][]{holoien17a,holoien17b,holoien17c,BROWN19}. This incompleteness persists through at least the late 2010s and was most severe before the year $\sim$ 2000 when major efforts to systematically discover SNe gained momentum \citep[e.g.,][]{LI11}.

We make use of the following information for each entry in the catalog: SN name, discovery date, host galaxy name, right ascension, declination, SN type, and $z$. We use a version of the OSC that contains $\sim$68,000 entries with discovery dates through February 2020.

\subsubsection{Selections to the Open Supernova Catalog}
\label{sec:osc cuts}

We apply the following criteria to the OSC to select a set of recent SNe that can be placed within individual galaxies. These criteria guarantee that our targets are likely to be true SNe, have occurred recently enough that they are unlikely to have strongly affected their environments, and can be placed within our galaxies. The cost of applying these criteria is that we may exclude some genuine recent SNe because they lack key data required for this study.

\begin{enumerate}

\item \textbf{Discovery date:} We remove entries without a discovery date. The main effect of this cut is to exclude supernova remnants (SNRs). SNRs have uncertain observational selection functions and have been catalogued in a more limited set of galaxies. Most importantly, because SNRs often occurred $\sim 10^4{-}10^5$~yr in the past, their explosive origins are unknown, and they have already exerted a significant effect on their local environment. Including them would therefore complicate the interpretation of our measurements. With this selection, 4,501 OSC entries are removed. While SNRs may only be a subset of these entries (albeit a large subset), removing those without a discovery date ensures that we are working with well-recorded SNe. Many SNRs are also naturally filtered out by excluding both the Milky Way and M31 from this study. Once we obtained our final sample, we manually checked each SN to ensure no SNRs leaked into our sample.

We also explored the effect of excluding SNe based on discovery date to avoid the cases where WISE and GALEX measure emission from the SNe themselves. To address similar concerns, \citet{anderson12} excluded SNe Ib/c and II that were discovered within 1~year and 1.5~years of galaxy imaging, respectively. In our data, there appears to be no statistically significant difference between a sample that excludes SNe based on these same criteria and a sample that includes them. Therefore, we do not limit our sample size based on when SNe were discovered in relation to galaxy imaging.

\item \textbf{Supernova types:} We remove all entries that do not correspond to confirmed SN events. Many of these entries are not given a type; that is, they have blank entries in the SN type column. Others are SN candidates or have ambiguous types (e.g., they are labeled ``Ia?").

In the end, we only keep entries designated as SNe Ia, II, II~P, II~L, IIb, Ib, Ic, and Ib/c. For analysis purposes, we combine SNe II, II~P, and II~L together under the label ``SNe II.'' We exclude SNe IIn entirely from this analysis because of the findings of \citet{anderson12}, which report statistically significant differences between SNe IIn and the general SNe II population (which further suggests that SNe IIn do not arise from the most massive stars, contradicting current theory). Due to a low number of SNe IIn in our sample, we could not perform a robust statistical analysis to support or contradict the results of \citet{anderson12}. Thus, we simply excluded them from the overall sample.

We then group SNe Ib, Ic, and those already labeled ``Ib/c" together as ``SNe Ib/c.'' We also include SNe IIb, which are hydrogen poor, in this category because they are observationally (and likely physically, since they are also stripped-envelope SNe) more related to SNe Ib and Ic (which are free of hydrogen and helium) rather than SNe II \citep{taddia18}. This leaves us with three categories of SNe Ia, II, and Ib/c, which we refer to throughout this paper.

\item \textbf{Right ascension and declination:} We consider only entries that include a right ascension (RA) and declination (Dec.) for the SNe. We require this information to identify the location of the SNe in their host galaxies. For a SN with multiple values of RA and Dec., we adopt the median of each coordinate as its location, as the variation in position is usually small relative to the galaxy size.
\end{enumerate}

After imposing these additional requirements, we retain $\sim$28\% of the OSC ($18,672$ entries), which corresponds to events confidently identified as SNe with clear discovery dates, positions, and types. We refer to this pared version of the OSC as our ``clean" catalog.

\subsubsection{Cross-Matching with the WISE and GALEX Maps} \label{sec:cross-match}

Next, we identify the members of this clean version of the OSC that lie within galaxies in the z0MGS atlas. To do this, we cross-match the locations of the SNe in the clean catalog with the GALEX and WISE maps at fixed angular resolutions (7.5\arcsec\ and 15\arcsec). We use a direct positional cross-match rather than, e.g., cross-matching on host name because not all SNe in the clean catalog have recorded hosts. We work with the original maps, i.e., before convolving to fixed physical resolution, because not all $z$0MGS galaxies can be convolved to physical resolutions $\leq 2$~kpc. Running a positional cross-check on the original images captures the largest possible set of SNe. This serves as the parent sample for our analysis.

\begin{figure*}[]
\centering
\includegraphics[width=.8\textwidth]{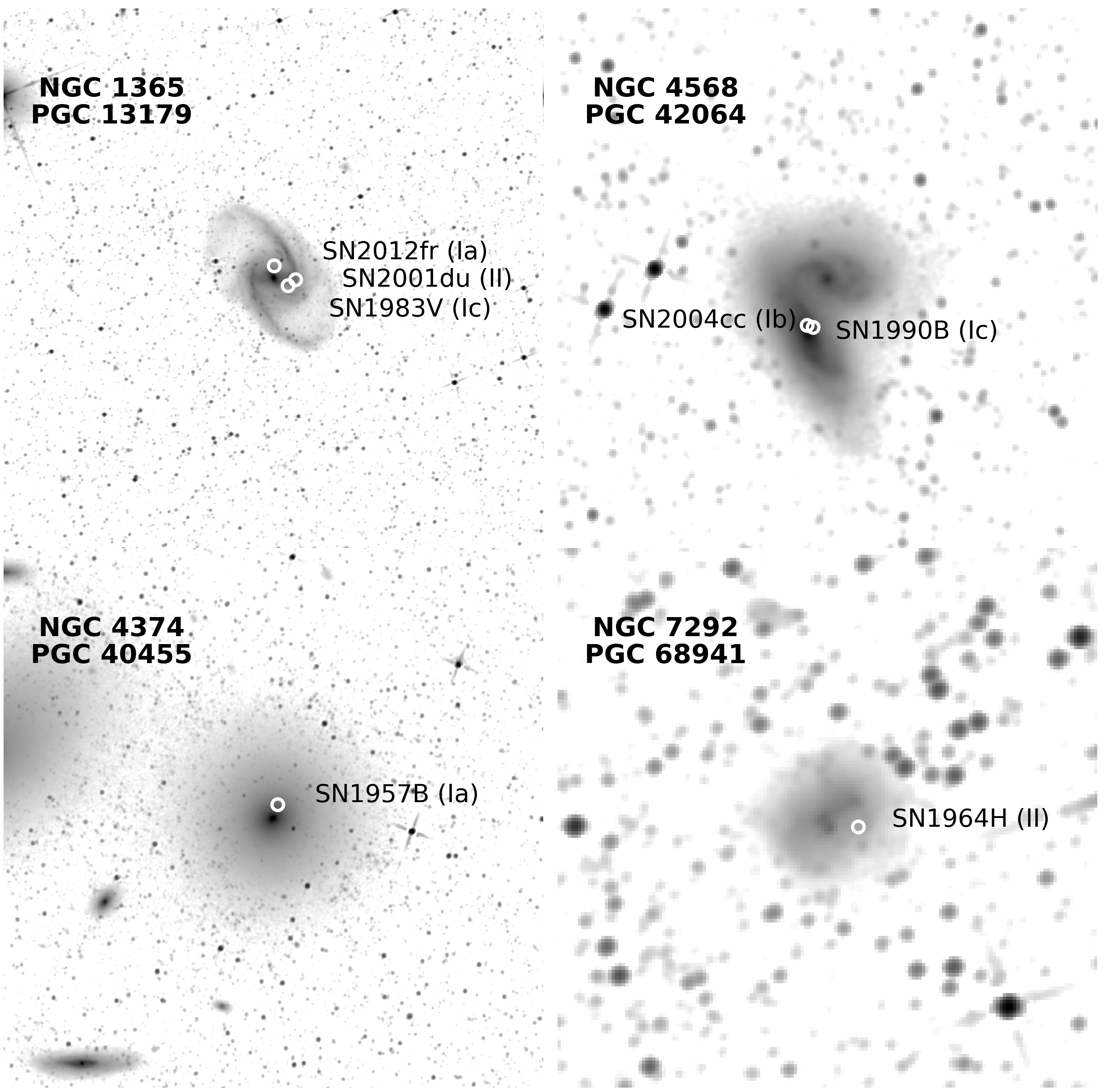}
\caption{\textbf{Illustration of SN placement into WISE Band 1 $3.4~\mu$m atlas images.} SN locations in W1 $3.4~\mu$m emission maps of four galaxies at 7.5\arcsec\ resolution. From top left to bottom right: NGC 1365 (PGC 13179); NGC 4568 (PGC 42064); NGC 4374 (PGC 40455); and NGC 7292 (PGC 68941). SN locations are marked with a white circle, with each SN labeled with its name and type. These images have not yet been processed with the masking and interpolation described in the text.}
\label{fig:locate_sne}
\end{figure*}

To carry out the cross-match, we test whether the adopted RA and Dec.~of each SN in the clean sample lies within the original $7.5\arcsec$ resolution W1 image for each galaxy. We adopt this approach because $7.5\arcsec$ W1 images exist for all galaxies in the $z$0MGS atlas. For SNe that lie within an image, we locate the pixel that corresponds to their location. Figure \ref{fig:locate_sne} shows an example of placing SNe in their host galaxy 7.5\arcsec\ ~W1 maps.

Each $z$0MGS image nominally corresponds to a single galaxy. In practice, the images can contain multiple galaxies, reflecting interactions, close pairs or groups, or simply chance superposition (e.g., see Figure \ref{fig:locate_sne}). Therefore, after this initial cross-matching, we must confirm that the SNe placed in an image are indeed associated with the nominal target of the image. To confirm the association between a SN and a galaxy, we take the following steps:

\begin{enumerate}

\item \textbf{Checking galactocentric radius:} For each atlas image that contains a SN, we construct an ellipse centered on the nominal target galaxy in the image. The ellipse has major axis of $2~r_{25}$. Here, $r_{25}$ is the 25$^{\rm th}$~mag~arcsec$^{-2}$ $B$ band isophotal radius, which is frequently adopted as the optical radius of a galaxy. We adopt $r_{25}$, inclination, and position angle mostly from the Lyon-Meudon Extragalactic Database \citep[{LEDA};][]{makarov14}, which in turn draws heavily on \citet{devaucouleurs91}. We require that SNe lie within this ellipse to be considered associated with that galaxy. This also means excluding SNe of host galaxies that are missing data required to construct this ellipse (i.e.,~$r_{25}$, inclination, and position angle).

\item \textbf{Cross-matching based on host name:} When the OSC provides a host name, we cross-match it with the name of the nominal target of the atlas, resolving any aliases using LEDA and the NASA Extragalactic Database\footnote{The NASA/IPAC Extragalactic Database (NED) is funded by the National Aeronautics and Space Administration and operated by the California Institute of Technology. \url{https://ned.ipac.caltech.edu}.}. In the case that the host name provided by the OSC matches that of the target galaxy in the image, we confirm the match.
    
\item \textbf{Checking by eye:} Some SNe were located inside $2~r_{25}$ of more than one galaxy, e.g., in cases of interacting galaxy pairs or galaxy groups. In these cases, we manually check W1 maps and assign the SNe to the galaxy that appeared most appropriate. We also carried out this visual quality check in the case of SNe without a host name in the OSC, confirming the association with the galaxy in the image.

\item \textbf{Checking redshift, $z$:} To ensure that SNe in our sample are not associated with background galaxies, we compare the redshifts of the SNe from the OSC with the redshifts of the associated galaxies in the atlas. We adopt the galaxies' recessional velocities from LEDA. If multiple redshifts are recorded for a SN, we take the median. If the difference between a SN's redshift and a galaxy's redshift is $\Delta z \lesssim 0.002$, then we accept the association of the SN with the galaxy. This $\Delta z$ corresponds to a velocity offset $\Delta V$ of $\approx 600$~km~s$^{-1}$ between the SN and the systemic velocity of the galaxy. This corresponds to approximately twice the maximum rotational velocity observed for spiral galaxies \citep[e.g.,][]{deblok08}.

In a small number of cases, an entry in the clean version of the OSC has neither a host name nor a redshift. There are only 12 SNe missing these data, and we exclude these cases from our sample because we cannot convincingly verify their association with the host galaxy.

\item \textbf{Excluding highly-inclined galaxies:} We exclude SNe in host galaxies that have inclinations $i\geq60^\circ$. Such heavily-inclined galaxies make it difficult to precisely estimate the galactocentric radius or local conditions associated with a SN.

\end{enumerate}

If the association of a SN with its host galaxy cannot be confirmed through steps 1-4, or is eliminated by step 5, then that SN is excluded from the sample. At this point, we identify $668$ total SNe in the OSC ($\sim$1\% of the original OSC) that are confidently associated with moderately-inclined galaxies in the $z$0MGS atlas.

The SN sample is further sub-divided based on physical resolution, as some $z$0MGS maps cannot be convolved to a resolution $\leq 2$~kpc, and wavelength coverage, as WISE covered more area of the sky than GALEX. Table~\ref{tab:sampsize} breaks down our final working sample, reporting the number of SNe per type available for each band at each physical resolution. We focus our analysis on the 2-kpc sample because it yields a much larger sample of SNe compared to the 0.5-kpc and 1-kpc resolution maps (see Table~\ref{tab:snsamp} in the Appendix). The 2-kpc resolution should still be small enough to capture the structure of the galaxy and galactic environment where the SNe occur. For reference, $7.5$\arcsec\ corresponds to 2~kpc at $d \approx 55$~Mpc, while $15$\arcsec\ corresponds to 2~kpc at $d \approx 27.5$~Mpc.

Table \ref{tab:galaxysamp} lists the number of galaxies that host SNe at each band and resolution. The 2-kpc sample retains the largest number of galaxies at 359, with 68 elliptical galaxies ($T < 2$), 245 spiral galaxies ($2 \leq T \leq 6$), 44 irregular/dwarf galaxies ($T > 6$), and 2 unspecified. The list of all 472 SNe in the 2-kpc sample can be found in Table~\ref{tab:snsamp} in the Appendix.


\begin{deluxetable}{c l c c c c c c}
\tabletypesize{\scriptsize}
\tablecaption{\label{tab:sampsize} Supernova Sample Size by Band, Type, and Resolution}
\tablewidth{0pt}
\tablehead{
\colhead{} & \colhead{}  & \multicolumn{3}{c}{Total per SN type}  & \multicolumn{3}{c}{Total per band} \\
\colhead{Band} & \colhead{SN Type} & \colhead{0.5~kpc} & \colhead{1~kpc} & \colhead{2~kpc} & \colhead{0.5~kpc} & \colhead{1~kpc} & \colhead{2~kpc}
}
\startdata
\multirow{3}{*}{\shortstack{W1 \\ (3.4~$\mu$m)}} &    SNe Ia  & 11  & 62 & 142   &  & & \\
{}  & SNe II  & 36  & 105 & 222  &  60 & 218 & 472 \\
{} &    SNe Ib/c & 13  & 51 &  108   &  & & \\
\hline
\multirow{3}{*}{\shortstack{W2 \\ (4.5~$\mu$m)}} &    SNe Ia  & 11 & 62 & 142   & & & \\
{} & SNe II  & 36 & 105 & 222   & 60 & 218 & 472 \\
{} &    SNe Ib/c & 13  & 51 & 108   & & & \\
\hline
\multirow{3}{*}{\shortstack{W3 \\ (12~$\mu$m)}} &    SNe Ia  & 11 & 62 & 142   & & & \\
{} & SNe II  & 36 & 105 & 222   & 60 & 218 & 472 \\
{} &    SNe Ib/c & 13  & 51 & 108   & & & \\
\hline
\multirow{3}{*}{\shortstack{W4 \\ (22~$\mu$m)}} &    SNe Ia  & 1  & 10 & 42    & & & \\
{} & SNe II  & 12 & 34 & 77    & 16 & 55 & 159\\
{} &    SNe Ib/c & 3  &  11 & 40   & & & \\
\hline
\multirow{3}{*}{\shortstack{NUV \\ (231~nm)}} &     SNe Ia  & 8  & 49 & 122  & & & \\
{} & SNe II  & 33 & 94 & 194  & 54 & 190 & 412\\
{} &     SNe Ib/c & 13  & 47 & 96   & & & \\
\hline
\multirow{3}{*}{\shortstack{FUV \\ (154~nm)}} &     SNe Ia  & 7  & 40 & 83   & & & \\
{} & SNe II  & 33 & 74 & 128  & 52 & 153 & 277\\
{} &     SNe Ib/c & 12  & 39 & 66   & & & \\
\hline
\multirow{3}{*}{\shortstack{$\Sigma_{\rm SFR}$ \\ (NUV$+$W3)}} &     SNe Ia  & 8  & 49 & 122    & & & \\
{} & SNe II  & 33 & 94 & 194   & 54 & 190 & 412 \\
{} &     SNe Ib/c & 13  & 47  & 96  & & & \\
\hline
\multirow{3}{*}{\shortstack{$\Sigma_{\rm SFR}$ \\ (FUV$+$W4)}} &     SNe Ia  & 1  & 7 & 35    & & & \\
{} & SNe II  & 12 & 32 & 69   & 16 & 50 & 140\\
{} &     SNe Ib/c & 3  & 11  & 36   & & & \\
\enddata
\tablecomments{SN sample size used in our analysis after confirming the association with the host galaxy and removing high-inclination ($i > 60^{\circ}$) targets from the sample. See details in Section \ref{sec:supernovae}.}
\end{deluxetable}

\begin{deluxetable}{l c c c}
\tablewidth{1.0\columnwidth} 
\tablecaption{\label{tab:galaxysamp} Host Galaxy Sample Size by Band and Physical Resolution.}
\tablehead{
\colhead{Band} & \colhead{0.5~kpc} & \colhead{1~kpc} & \colhead{2~kpc}
}
\startdata
W1 (3.4~$\mu$m) & 37  & 150 & 359 \\
W2 (4.5~$\mu$m) & 37  & 150 & 359 \\
W3 (12~$\mu$m) & 37  & 150 & 359 \\
W4 (22~$\mu$m) & 7   & 32  & 102 \\
NUV (231~nm) & 32  & 125 & 304 \\
FUV (154~nm) & 30  & 98  & 189 \\
$\Sigma_{\rm SFR}$(FUV+W4) & 7   & 28  & 86  \\
$\Sigma_{\rm SFR}$(NUV+W3) & 32   & 125  & 304  \\
\hline 
Maximum & 37 & 150 & 359 \\
\enddata
\tablecomments{Host galaxy sample size per band and per physical resolution after confirming the association of SNe and removing high-inclination targets. The sample size for $\Sigma_{\rm SFR}$(FUV$+$W4) is the overlap between the sample sizes of the FUV and W4 bands. The sample size for $\Sigma_{\rm SFR}$(NUV$+$W3) is the overlap between the sample sizes of the NUV and W3 bands. See details in Sections~\ref{sec:galaxies} and \ref{sec:supernovae}.}
\end{deluxetable}

\section{Supernova Environments in $\Sigma_{\rm SFR}$-$\Sigma_\star$ Space
} 
\label{sec:sfms}

\begin{figure*}
\centering
\includegraphics[width=0.75\textwidth]{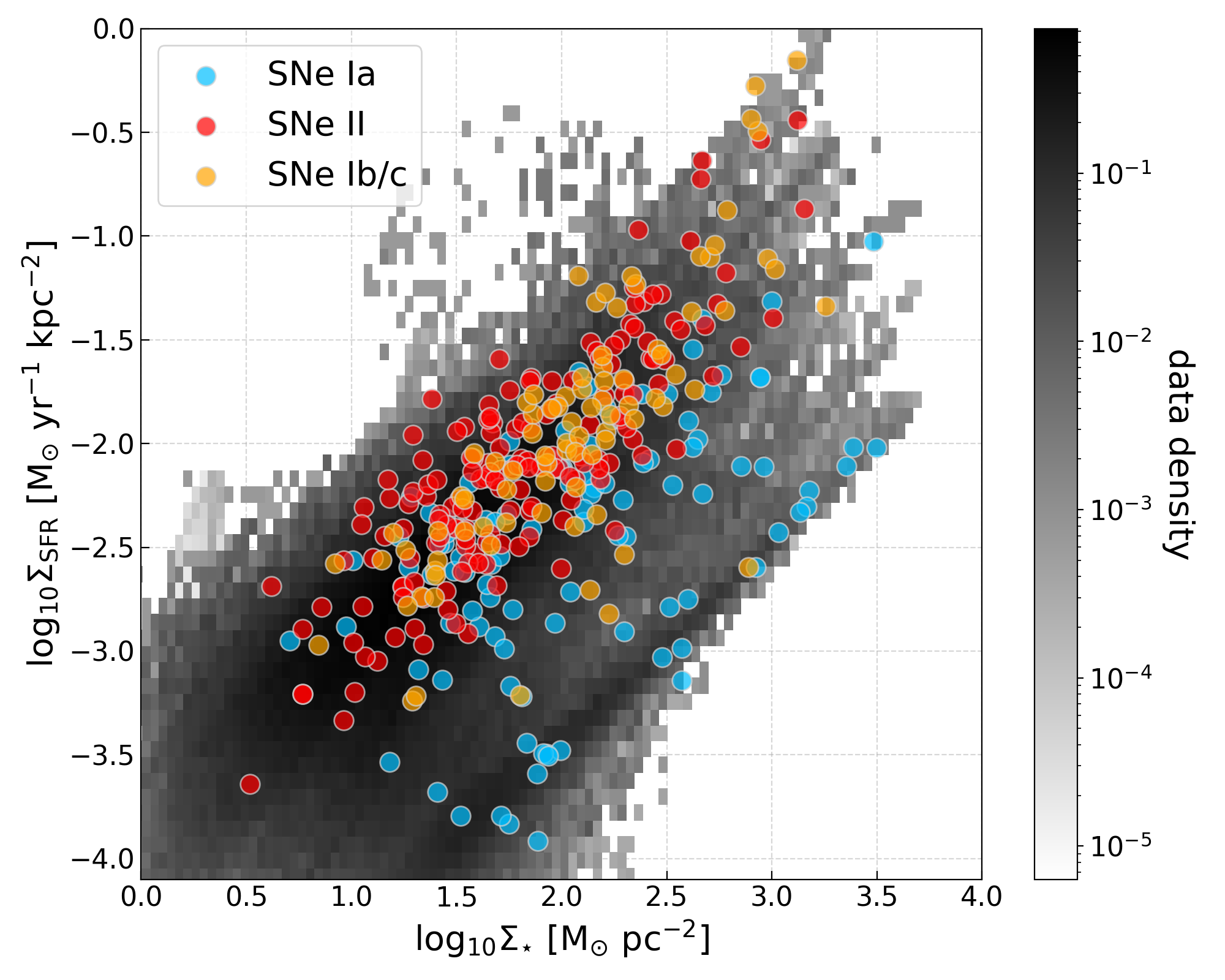}
\includegraphics[width=1.0\textwidth]{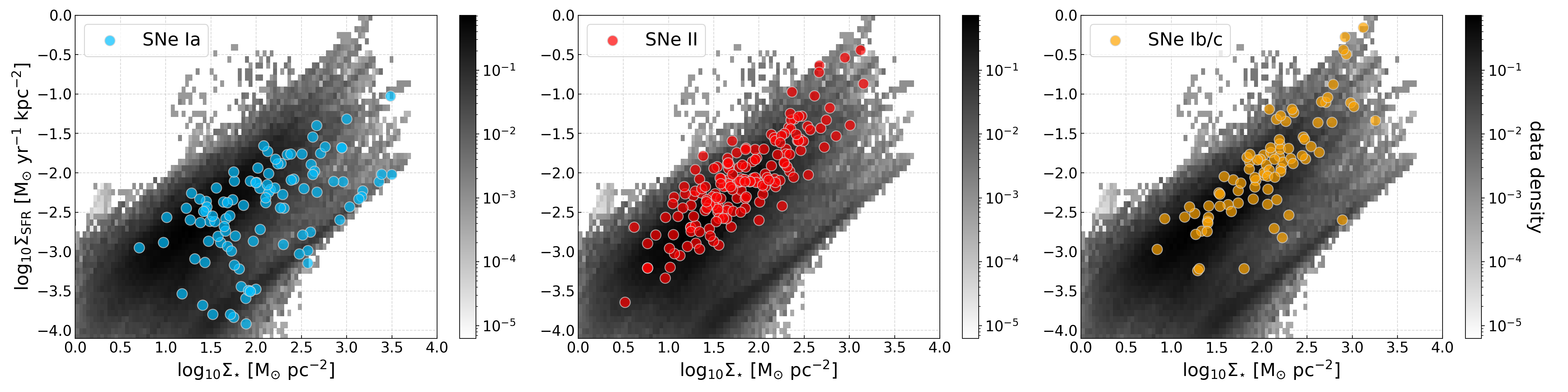}
\caption{\textbf{The environments of SNe in the $\Sigma_{\rm SFR}{-}\Sigma_\star$ plane.} Local, 2-kpc resolution values of $\Sigma_{\star}$ and $\Sigma_{\rm SFR}$ at the sites of SNe appear as colored points: SNe Ia in blue, SNe II in red, and SNe Ib/c in orange. In grayscale (i.e., the ``data density" colorbar), we show the weighted distribution of $\Sigma_{\rm SFR}$ and $\Sigma_\star$ for all regions within $r_{25}$ of all host galaxies in the $z0$MGS sample. In detail, each pixel count is weighted by $1/N$, where $N$ is the total number of pixels inside each galaxy. Top: all SN types are plotted in one panel. Bottom: each SN type is plotted separately.}
\label{fig:sfr_vs_star}
\end{figure*}

We measure IR and UV emission from the site of each SN in our sample. Then, following Section \ref{sec:sfr}, we use our measurements to estimate the surface densities of stellar mass and recent star formation rate near each SN. For this analysis, we trace star formation using a combination of NUV and W3, $\Sigma_{\rm SFR}$(NUV$+$W3). This allows us to retain the largest possible sample of SNe at the expense of some likely contamination of $\Sigma_{\rm SFR}$ at low values of $\Sigma_{\rm SFR}/\Sigma_\star$ \citep[see Section \ref{sec:sfr} and][]{leroy19}. We estimate $\Sigma_\star$ from W1 emission, applying the mass-to-light ratio estimated for each galaxy by \citet{leroy19}. We correct both axes for the effects of inclination.

In Figure \ref{fig:sfr_vs_star}, we visualize the results of these measurements by showing the locations of each SN in $\Sigma_{\rm SFR}{-}\Sigma_\star$ space. These diagrams show the 2-kpc resolution SFR per unit area, $\Sigma_{\rm SFR}$, as a function of stellar mass per unit area, $\Sigma_\star$. We plot measurements for the 2-kpc resolution environment associated with each SN as colored points, with the color reflecting the SN type.

In grayscale, we show the density of points from \textit{all} regions, with or without SNe, that have a galactocentric radius $r_{\rm gal} < r_{\rm 25}$ in our host galaxies. Less formally, the gray image shows how the full area in our sample of SN host galaxies is distributed across this space. We visualize the data density as a grayscale color on a logarithmic stretch. Thus, the grayscale represents all locations where a SN {\it could} have exploded in our SN host galaxies. The colored points show where they {\it did} explode.

To construct the grayscale comparison image, we create a grid in $\log_{10} \Sigma_{\rm SFR}{-}\log_{10} \Sigma_\star$ space with grid cells $0.04 \times 0.08$~dex in size. Our galaxy maps are convolved to fixed physical resolution; however, the pixel size is still at fixed angular resolution. To make a 2D-histogram with equal-weight in physical resolution, we assign a weight of $N_{\rm pix}^{-1}$, where $N_{\rm pix}$ is the total number of pixels in each host galaxy inside $r_{25}$.

The diagonal strip visible in both the SNe (colored points) and the control (grayscale) image at high $\Sigma_{\rm SFR}$ illustrates the ``resolved star forming main sequence,'' (RSFMS). The RSFMS reflects a well-defined relationship between $\Sigma_{\rm SFR}$ and $\Sigma_\star$ exhibited by the star-forming parts of late-type galaxies \citep{cano-diaz16,medling18}. This relationship mirrors the ``main sequence of star forming galaxies,'' or the scaling between integrated SFR and stellar mass observed for unresolved galaxies \citep[e.g., see][]{speagle14}. The regions showing lower $\Sigma_{\rm SFR}$ at fixed $\Sigma_\star$ correspond to early-type galaxies or to the bulges and quiescent regions in late-type galaxies.

The environments of SNe populate this diagram in a systematic way that largely reflects our expectations. CC SNe, i.e.,~SNe II and SNe Ib/c, lie almost exclusively along or above the RSFMS.\footnote{We note that 13 CC SNe appear in more quiescent regions of their galaxies with $\log_{10}\Sigma_{\rm SFR}/\Sigma_\star < -10.5$. None of these CC SNe were misclassified in the OSC nor labeled as ``Ca-rich'' events. A partial explanation is that 6 of these SNe at such low $\Sigma_{\rm SFR}/\Sigma_\star$ values occurred in early-type ($T < 2$) galaxies. These may be potential targets of interest for a future study, which would benefit from sharper resolutions.} This is consistent with these events stemming from the deaths of recently formed, massive stars; i.e., the CC SNe in Figure \ref{fig:sfr_vs_star} appear consistent with tracing $\Sigma_{\rm SFR}$ rather than $\Sigma_\star$. Furthermore, SNe Ib/c appear concentrated to the high end of this sequence, occurring mostly in regions of high $\Sigma_{\rm SFR}$ and high $\Sigma_\star$. The figure suggests that SNe Ib/c may be preferentially located in regions with the highest $\Sigma_{\rm SFR}$ rather than distributed following the overall $\Sigma_{\rm SFR}$ distribution.

Compared to CC SNe, SNe Ia show a much wider distribution of $\Sigma_{\rm SFR}$ at fixed $\Sigma_\star$. As a result, they appear more spread out in the diagram. We do find SNe Ia located along the RSFMS, but also in the more quiescent, low $\Sigma_{\rm SFR}$-high $\Sigma_\star$ regions found in early-type galaxies or bulges. Overall, this appears consistent with the idea that SNe Ia track the overall distribution of stellar mass, not only the locations of recent star formation.

\begin{figure*}
\centering
\includegraphics[width=0.48\textwidth]{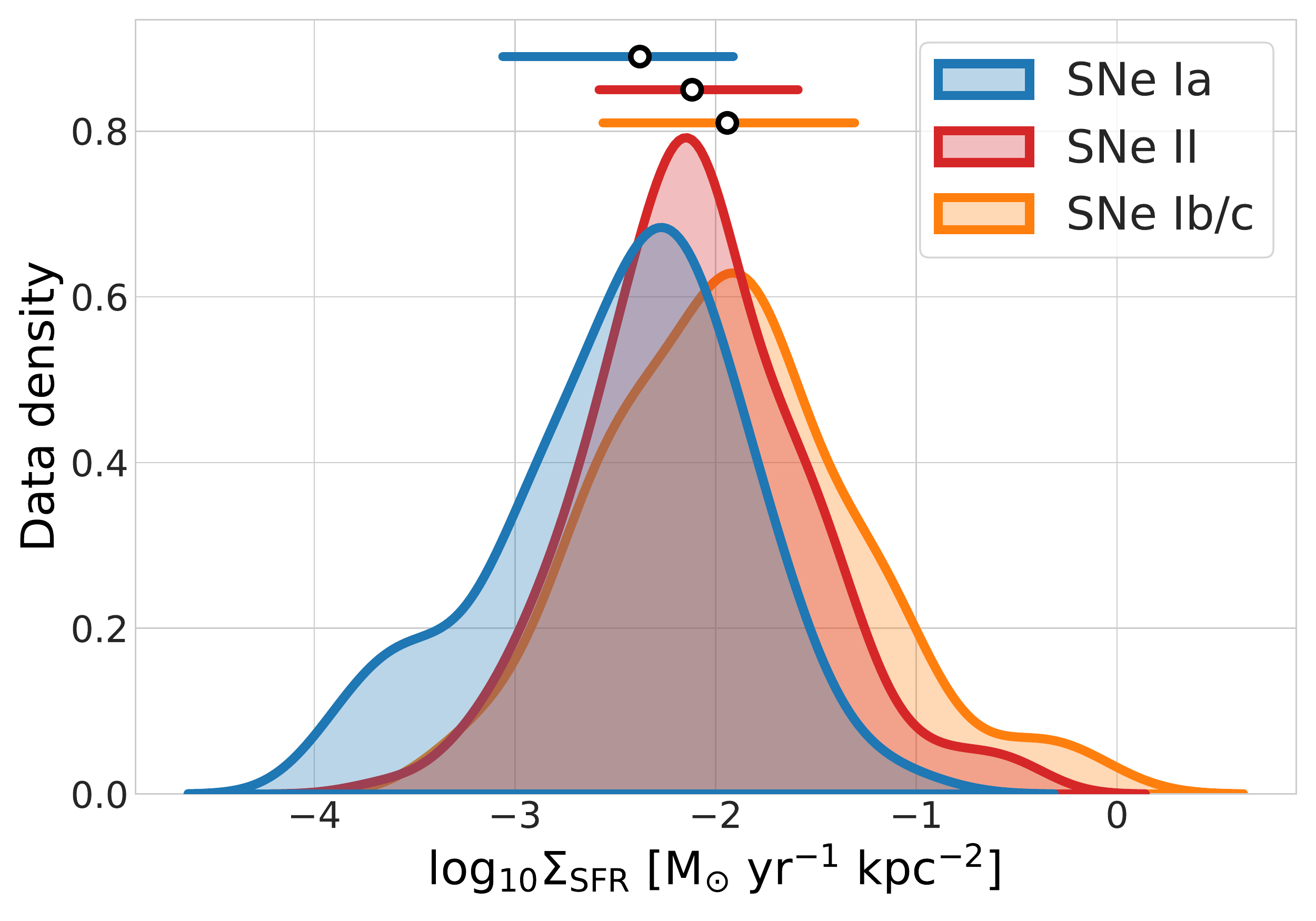}
\includegraphics[width=0.48\textwidth]{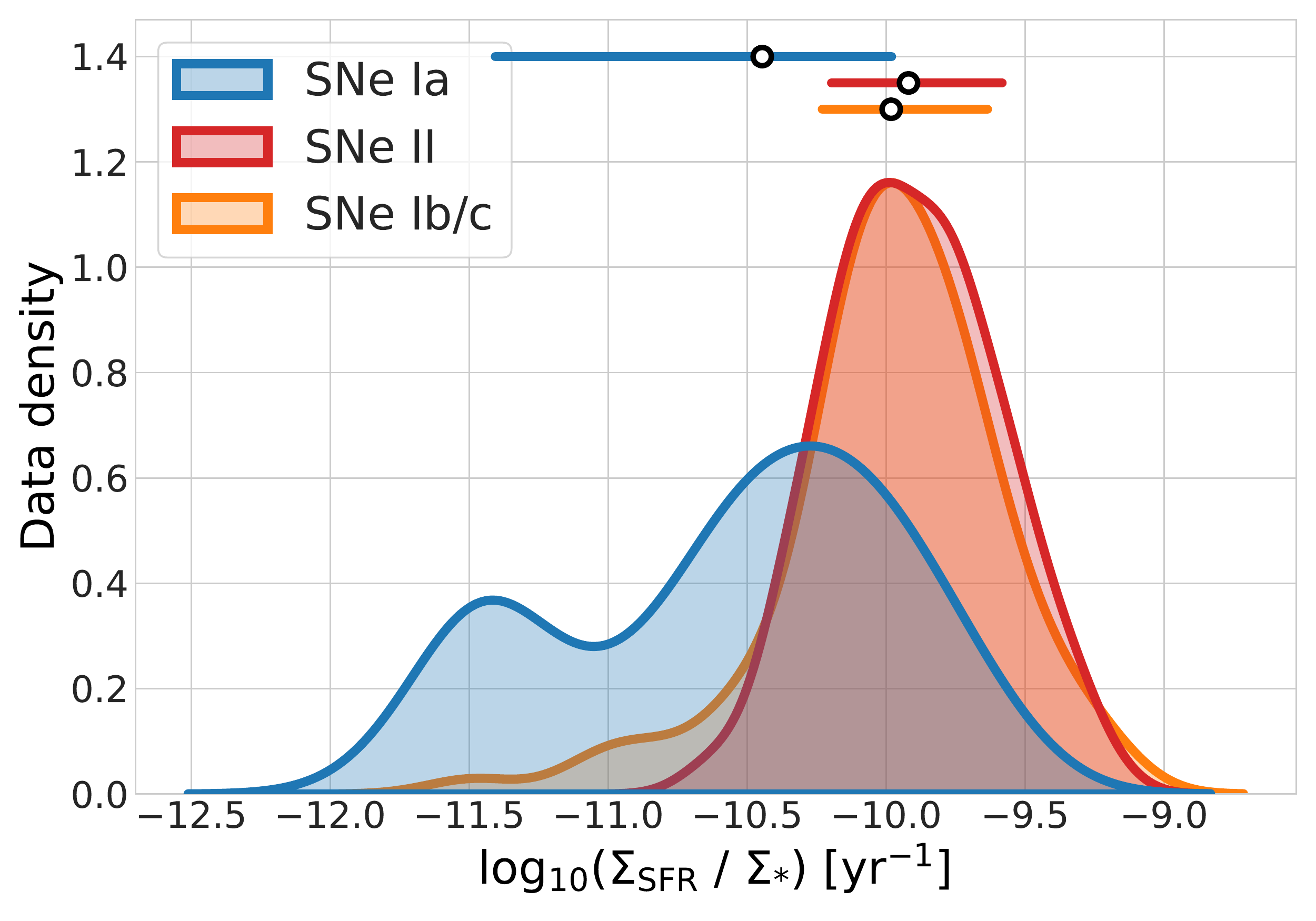}
\caption{\textbf{Kernel density estimation (KDE) plots for all SN types as a function of $\log_{10}\Sigma_{\rm SFR}$ and $\log_{10}(\Sigma_{\rm SFR}$/$\Sigma_{\star})$}. The {\bf left panel} shows the KDE plots for SNe Ia, II, and Ib/c as a function of $\log_{10}\Sigma_{\rm SFR}$. The horizontal bars span the $16^{\rm th}-84^{\rm th}$ percentiles of each distribution, with the circle marking the $50^{\rm th}$ percentile. These $16-50-84^{\rm th}$ percentiles are, respectively, $-3.06, -2.38, -1.91$ for SNe Ia; $-2.58, -2.12, -1.59$ for SNe II; and $-2.56, -1.94, -1.31$ for SNe Ib/c. These distributions show that SNe Ib/c are shifted to higher values of $\log_{10}\Sigma_{\rm SFR}$, while SNe Ia are shifted to the lowest values. The {\bf right panel} displays the KDE plots for these SN types as a function of $\log_{10}(\Sigma_{\rm SFR}$/$\Sigma_{\star})$. The horizontal bars again span the $16^{\rm th}-84^{\rm th}$ percentiles of each distribution, with the median marked by a circle. These $16-50-84^{\rm th}$ percentiles are, respectively, $-11.41, -10.45, -9.98$ for SNe Ia; $-10.2, -9.92, -9.58$ for SNe II; and $-10.23, -9.98, -9.64$ for SNe Ib/c. We see that SNe Ia have a much broader distribution and are more concentrated to lower values of $\log_{10}(\Sigma_{\rm SFR}$/$\Sigma_{\star})$. The SN II and Ib/c distributions are quite similar, with the SN Ib/c distribution appearing slightly narrower (i.e.,~more concentrated) than the SN II distribution, and the SN II median shifted to a slightly higher $\log_{10}(\Sigma_{\rm SFR}$/$\Sigma_{\star})$ than SNe Ib/c.}
\label{fig:kde}
\end{figure*}

In Figure \ref{fig:kde}, we directly visualize the distributions of star formation rate surface density, $\Sigma_{\rm SFR}$, and local specific star formation rate, $\Sigma_{\rm SFR}/\Sigma_\star$, for SN environments. The figure shows data density distributions for the $\log_{10}(\Sigma_{\rm SFR})$ and $\log_{10}(\Sigma_{\rm SFR}/\Sigma_{\star})$ associated with each SN type. We estimate the distributions via kernel density estimation \citep[KDE;][]{scott92}. The figures include horizontal bars to span the $16^{\rm th}-84^{\rm th}$ percentiles of each distribution, with circles marking the medians.

We conduct a \citet{student} t-test to estimate the probability of generating these different median values from the same parent distribution (see Table \ref{tab:t-test} in Appendix \ref{sec:t-test}). We use the resulting $p-$values to evaluate the significance of the differences in Figure \ref{fig:kde}. We will reject the null hypothesis that the median values could be generated from the same parent distribution when the $p-$value $< 5\%$.

Figure \ref{fig:kde} supports our qualitative observations from Figure~\ref{fig:sfr_vs_star}. The left panel shows that SN Ib/c distribution is concentrated to higher values of $\Sigma_{\rm SFR}$ compared to either SNe II or SNe Ia, with SNe II occupying an intermediate regime. The order of the median $\log_{10} \Sigma_{\rm SFR}$ values is: SNe Ia ($-2.38$) $<$ SNe II ($-2.12$) $<$ SNe Ib/c ($-1.94$). Furthermore, $p-$values of the \citet{student} t-test are all $< 5\%$, indicating that the median $\log_{10} \Sigma_{\rm SFR}$ of these SN distributions are statistically different from each other.

The right panel of Figure~\ref{fig:kde} shows the distributions of $\log_{10}(\Sigma_{\rm SFR}/\Sigma_{\star})$ for each SN type. The SN Ia, II, and Ib/c distributions of $\log_{10}(\Sigma_{\rm SFR}/\Sigma_{\star})$ have the following median values: SNe Ia ($-10.45$) $<$ SNe Ib/c ($-9.98$) $<$ SNe II ($-9.92$). The median values for SNe II and Ib/c are similar, which supports the visual impression from Figure \ref{fig:sfr_vs_star} that the environments of SNe II and Ib/c are distributed similarly relative to the RSFMS. Furthermore, the $p-$value between the median of the SN II and Ib/c distributions is $7.8\%$, thus suggesting that these two median values could have resulted from the same parent distribution. We do note that when SNe IIb is included in the SNe II category rather than the SNe Ib/c category, the $p-$value $=42\%$, suggesting that SNe IIb have lower $\log_{10}(\Sigma_{\rm SFR}/\Sigma_{\star})$ values than SNe Ib/c.

The environments of SNe Ia, by contrast, show a broad distribution that extends to lower values of $\log_{10}(\Sigma_{\rm SFR}/\Sigma_{\star})$ compared to those found for CC SNe. Of particular note, the SN Ia distribution extends below the RSFMS to the resolved equivalent of the ``green valley'' \citep[i.e., intermediate galaxies between the star-forming main-sequence and quiescent galaxies;][]{martin07,salim07}, and includes a secondary peak at quiescent low values of $\Sigma_{\rm SFR}/\Sigma_{\star}$. As Figure \ref{fig:sfr_vs_star} shows, these regions have abundant stars but lack of recent star formation. SNe Ia still appear in these regions, consistent with many of these events arising from the older stellar population. Furthermore, the double-peak structure of the SN Ia distribution (blue curve), which is not present in the CC SN distributions, indicates that the host galaxy types (quiescent early types vs. star-forming late types) shape the distributions for SNe Ia, i.e. SNe Ia occur in elliptical galaxies (left peak) as well as in regions of spiral galaxies that lack star formation (right peak).

\begin{figure*}
\centering
\includegraphics[width=0.8\textwidth]{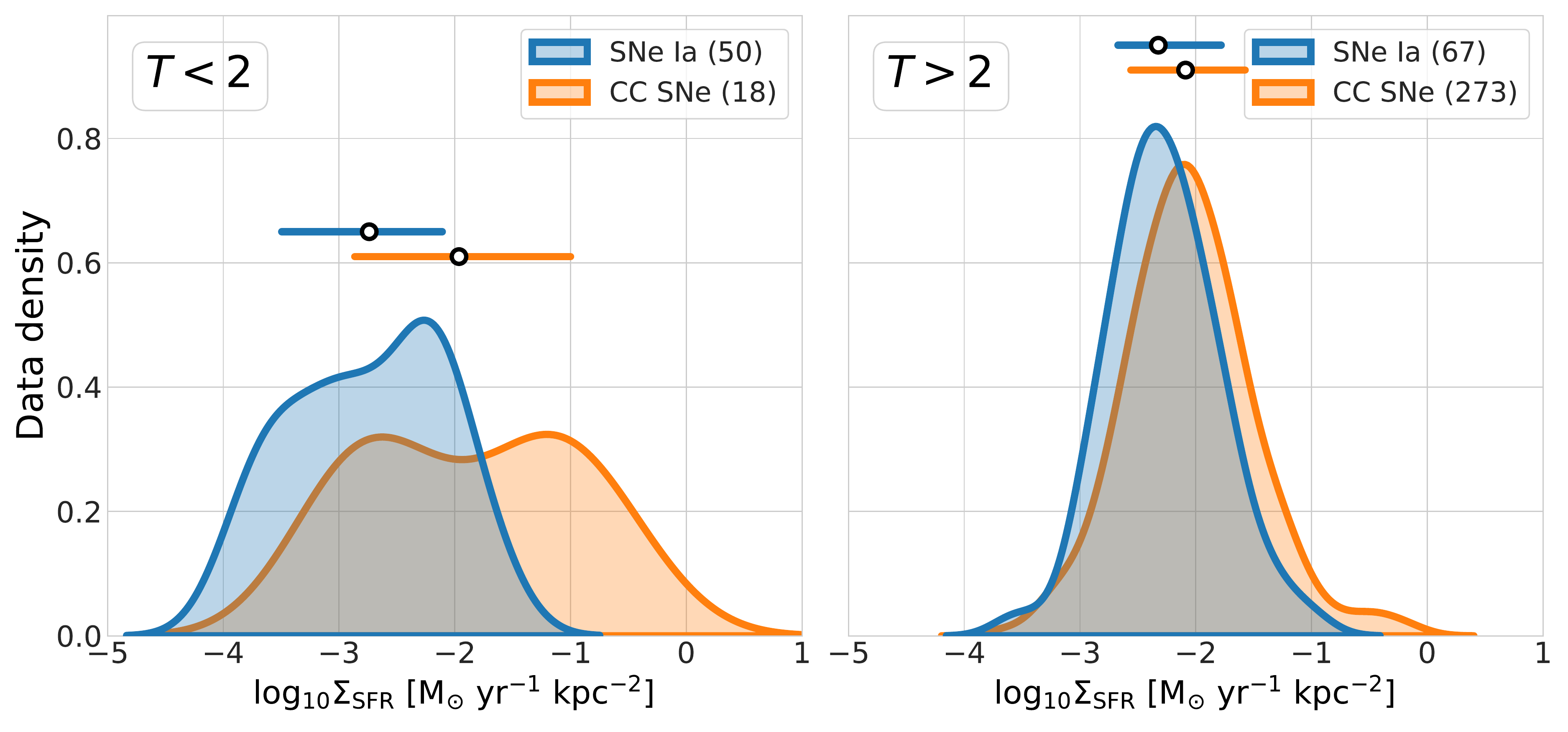}
\includegraphics[width=0.8\textwidth]{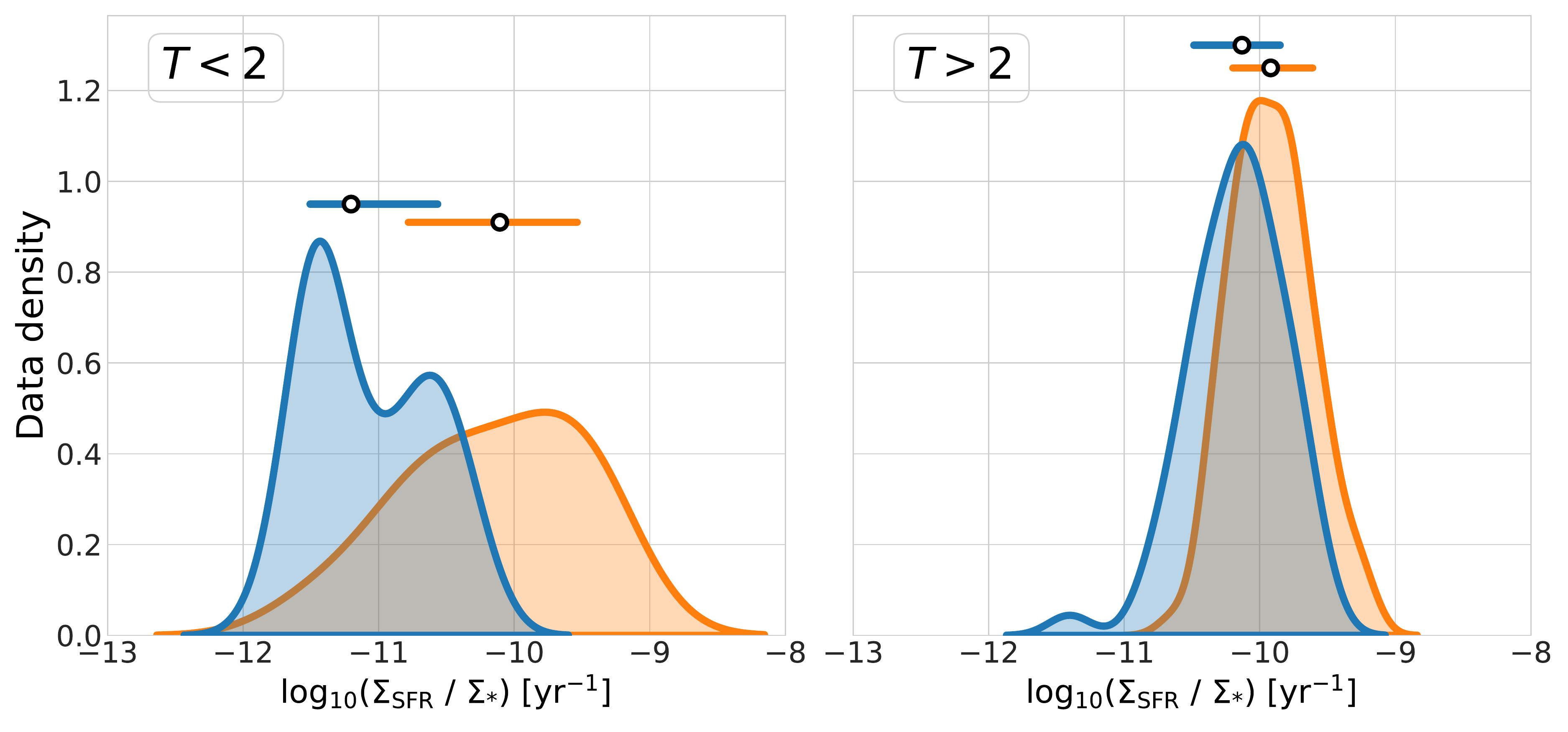}
\caption{{\textbf{Kernel Density Estimation (KDE) for $\log_{10}\Sigma_{\rm SFR}$ (top panels) and $\log_{10}(\Sigma_{\rm SFR}$/$\Sigma_{\star})$ (bottom panels), now separating the sample by morphology}. The figure resembles Figure \ref{fig:kde}, but now we group the SN sample into SNe Ia (blue) and CC SNe (orange) and separate SNe found in early-type ($T < 2$, left panels) and SNe found in late-type ($T > 2$, right panels) galaxies. The numbers in the parenthesis indicate the number of SNe. The horizontal bars span the $16^{\rm th}-84^{\rm th}$ percentiles of each distribution, with the circle marking the $50^{\rm th}$ percentile. The figure shows expected, strong differences between the $\Sigma_{\rm SFR}$ and $\Sigma_{\rm SFR}/\Sigma_star$ in early- and late-type galaxies. That is, the type of galaxy plays a strong role in setting the environments within the galaxy. Consistent with this and with previous results, a large fraction of our CC SNe are found in late-type galaxies, while our SN Ia sample is more evenly split by type. We also see that even within each morphological subsample, both $\Sigma_{\rm SFR}$ and $\Sigma_{\rm SFR}$/$\Sigma_{\star}$ vary between SNe Ia and CC SNe.}}
\label{fig:ia-cc-kde}
\end{figure*}

To illustrate this effect, we split the SN Ia and CC SN samples according to morphological type of their hosts, creating separate sub-samples for early-type ($T < 2$) and late-type ($T > 2$) galaxies. We simplify this analysis by combining all CC SNe (i.e., SNe II and SNe Ib/c together) and comparing them to SNe Ia. The number of SNe Ia spreads fairly between the sub-sample, with 50 SNe Ia in early-type galaxies and 67 in late-type galaxies. Meanwhile, the CC SNe are quite rare in early-type galaxies, as our sample includes only 18 CC SNe in early type galaxies compared to 273 SNe in late-types. Figure \ref{fig:ia-cc-kde} shows the resulting distributions for $\log_{10}(\Sigma_{\rm SFR})$ in the top panels and $\log_{10}(\Sigma_{\rm SFR}/\Sigma_{\star})$ in the bottom panels. The $16-50-84^{\rm th}$ percentiles of $\log_{10}\Sigma_{\rm SFR}$ are $-2.67,-2.32,-1.78$ for SNe Ia and $-2.56, -2.09, -1.57$ for CC SNe. The $16-50-84^{\rm th}$ percentiles of $\log_{10}(\Sigma_{\rm SFR}$/$\Sigma_{\star})$ are $-10.49, -10.13, -9.85$ for SNe Ia and $-10.2, -9.92, -9.61$ for CC SNe.

Figure \ref{fig:ia-cc-kde} shows that the difference between SN Ia and CC SN environments still exists \textit{within} the early-type and late-type subsamples. The Student's t-test confirms that this difference is statistically significant. That is, SNe Ia and CC SNe occur in different environments within the same galaxy type, so that even within late type galaxies, SNe Ia occur in more quiescent regions compared to CC SNe. The combined pictures from Figures \ref{fig:sfr_vs_star}, \ref{fig:kde}, and \ref{fig:ia-cc-kde} is that host galaxy plays a large, but not complete, role in setting what local environments are present in a galaxy, and that the SNe link to variations in both host galaxy properties and local environment within a galaxy in the qualitatively expected ways.

To summarize, Figures~\ref{fig:sfr_vs_star}, \ref{fig:kde}, and \ref{fig:ia-cc-kde} provide an overview of where different types of SNe explode in galaxies in terms of local SFR and local stellar mass surface densities. Our measurements largely conform to expectations that CC SNe explode in actively star-forming regions that have high $\Sigma_{\rm SFR}/\Sigma_\star$. We believe that this represents the first time that CC SNe have been shown to follow the resolved star-forming main sequence in galaxies, and indeed the first time that SN environments have been examined in this particular way. The analysis also supports the view that SNe Ib/c are preferentially found in even more active regions than SNe II, with the environments of SNe Ib/c displaced to somewhat higher $\Sigma_{\rm SFR}$ compared to SNe~II. By contrast, SNe Ia show a distribution that more closely tracks $\Sigma_\star$, exhibiting a wide range of $\Sigma_{\rm SFR}$ at fixed $\Sigma_\star$, again consistent with expectations that many of these explosions occur in the older stellar population, including early-type galaxies.

\section{Are Supernovae Radially Distributed like IR and UV Emission?} \label{sec:cdf}

In Section~\ref{sec:sfms}, we estimated $\Sigma_{\rm SFR}$ and $\Sigma_\star$ at the sites of SNe and compared these to the full distribution of environments in SN host galaxies in $\Sigma_{\rm SFR}{-}\Sigma_\star$ space. Here, we compare the radial distribution of SN locations within galaxies to the radial distributions of host galaxy IR and UV emission in the same galaxies. To our knowledge, this is the first large statistical comparison between SN locations and the distribution of $3.4{-}22~\mu$m emission in galaxies, as well as the first systematic comparison of SN locations to extinction-corrected $\Sigma_{\rm SFR}$ estimates. We also add a large, uniform comparison of SN location to the distribution of UV light traced by GALEX.

From a physical perspective, we test the hypotheses that 1.)~CC SNe follow the distribution of recent star formation because they result from young, massive stars, and 2.)~SNe Ia are distributed similar to the stellar mass in a galaxy because they originate from evolved stellar systems (see Section~$\ref{sec:intro}$). Both hypotheses appeared well-supported by the results in Section \ref{sec:sfms} and Figure~\ref{fig:sfr_vs_star}. Here, we expect these hypotheses to manifest as a close correspondence between the radial distributions of CC SNe and that of mid-IR emission (W3, W4), as well as a match between the location of CC SNe and the local $\Sigma_{\rm SFR}$(FUV$+$W4, NUV$+$W3). Meanwhile, we expect SNe Ia to be radially distributed in a similar way to near-IR emission (W1, W2), which trace the distribution of stellar mass.

\subsection{Method} 
\label{sec:method}

Only one or a handful of SNe explode in any given non-starburst, $z=0$ galaxy over the span of a century. Thus, any detailed comparison of SN locations to galaxy structure requires analyzing large samples of SNe spread across many galaxies. To do this, we need an approach that controls for variations in the internal structure of galaxies. In this section, we largely follow the method developed by \citet{james06} and \citet{anderson12, anderson15}. In this approach, we measure the intensity of light at the 2-kpc environment of each SN. We place this measurement within the context of the total distribution of light of that SN's host galaxy. We are then able to see if SN environments appear consistent with being drawn from the same distribution as the light of their host galaxies at any particular band. If a distribution of SNe and a distribution of host galaxy light differ at a significant level, then this comparison can provide evidence that the progenitor population for that type of SNe is not perfectly traced by that band.

In practice, this approach has three steps. First, we build the cumulative distribution function (CDF) of flux as a function of galactocentric radius for each host galaxy at each band. Then, we sample each CDF at the locations of known SNe and construct an aggregate distribution of CDF values at the location of SNe. Finally, we use this aggregated distribution of CDF values to test whether the SN locations show statistically significant deviations from the band used to construct the CDFs. We explain each step in greater detail in the next subsections.

\subsubsection{Building the Host Galaxy CDFs}

In the first step, we sort all of the pixels in each image of each host galaxy according to their galactocentric radius. Then, for each host galaxy, we construct a fractional CDF of flux in each band as a function of radius. This fractional CDF runs from 0 (at the galaxy center) to 1 (at a radius of $2~r_{25}$). In detail, we perform the following individual calculations:

\begin{enumerate}

\item \textbf{Record host galaxy intensity values:} For each galaxy, we measure the intensity in each pixel at a galactocentric radius less than $2~r_{25}$. During this step, we set all negative emission to zero. Negative pixels can occur in the maps due to noise fluctuations or, in rarer cases, artifacts related to convolution or background subtraction.

\item \textbf{Sort intensities by galactocentric radius:} For each galaxy, we sort the measured intensity values according to their distance from the galaxy center, from 0 to $2~r_{25}$.

\item \textbf{Normalize by the total integrated emission from the galaxy:} Next, we calculate the sum of all fluxes within $2~r_{25}$ for each galaxy. Then, we normalize all measured fluxes by this value. After this, each measured, normalized flux value corresponds to the fraction of the full galaxy flux at that band in that pixel.

\item \textbf{Build a normalized cumulative distribution function (CDF):} Finally, we combine these sorted, normalized measurements into a normalized radial CDF for each host galaxy. This curve measures the fraction of the total flux enclosed within each galactocentric radius.

\end{enumerate}

Figure \ref{fig:gal_cdf} is an example of one such CDF (which we often refer to as a ``host galaxy CDF''). We show the normalized radial CDF of W1 intensity at 2-kpc resolution in NGC 1365, which we construct from the masked, convolved version of the W1 image shown in Figure \ref{fig:locate_sne}. The $x-$axis corresponds to galactocentric radius, and the $y-$axis indicates the fraction of total flux enclosed within that radius. The black line in this figure shows the normalized CDF of W1 emission from the galaxy.

We repeat these steps for each $2$-kpc resolution WISE, GALEX, and $\Sigma_{\rm SFR}$ map that hosts a SN.

\subsubsection{Building the Distribution of Supernova CDF Values}

\begin{figure}[]
\centering
\includegraphics[width=.48\textwidth]{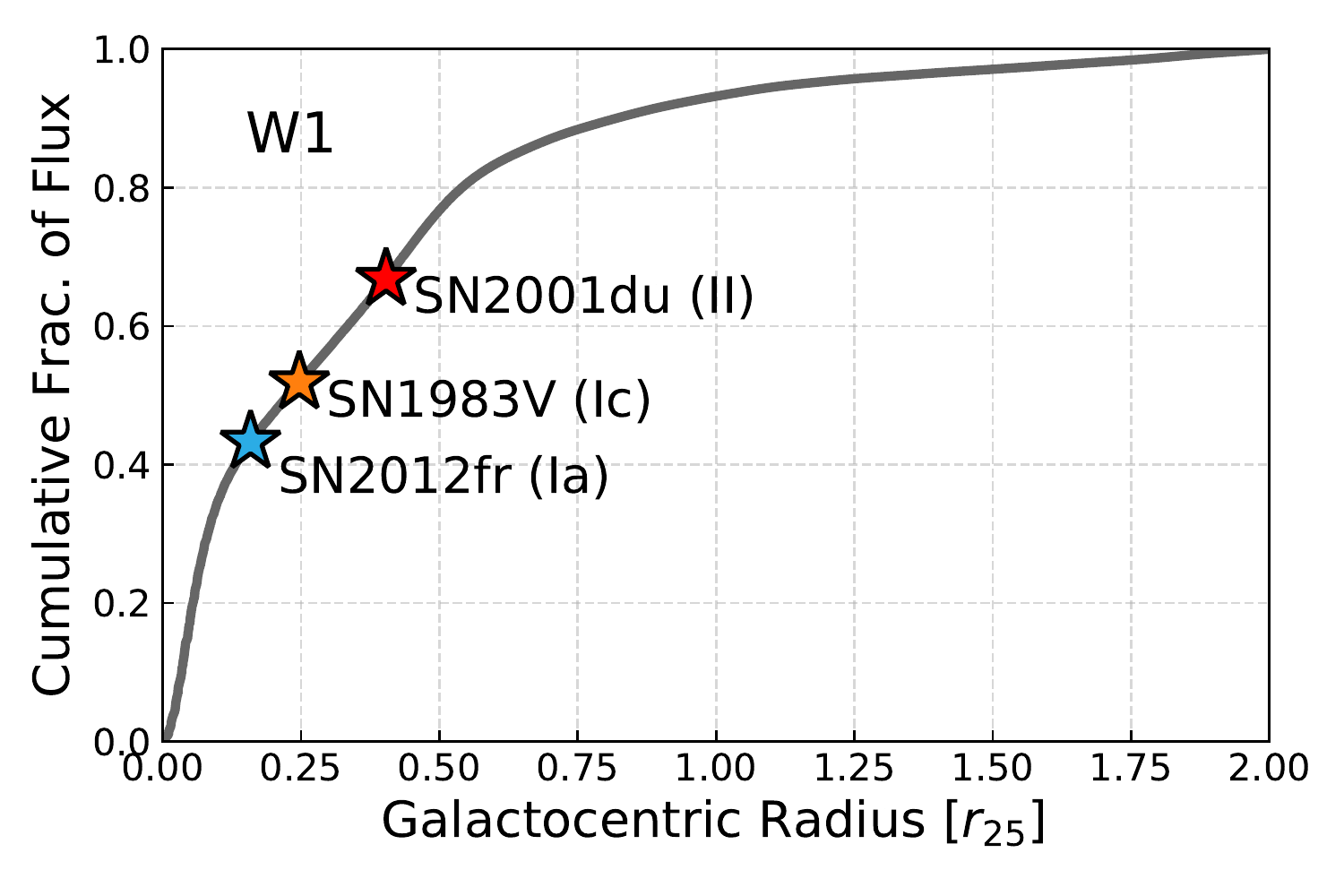}
\caption{\textbf{Locating SNe in the cumulative radial distribution of near-IR flux.} This figure shows the radial cumulative distribution function (CDF) of W1 ($3.4 \mu$m) flux in NGC 1365 (PGC 13179) out to $2~r_{25}$. The black curve records the fraction of the total W1 flux inside each galactocentric radius. The star symbols mark the galactocentric radius of the three recent SNe that occurred in this galaxy and were recorded in the OSC. We measure the W1 CDF value for each of these SNe by interpolating the host galaxy CDF (the black curve) to the galactocentric radius of SNe. We repeat this process for each SN in each band and use these collected CDF values to compare the distributions of SNe with respect to the IR and UV emission of host galaxies.}
\label{fig:gal_cdf}
\end{figure}

Next, we build a distribution of CDF values at the location of SNe in each band. To do this, we first calculate the galactocentric radius of each SN and record the value of the host galaxy CDF at that location. In other words, we measure the fraction of host galaxy flux within the galactocentric radius of each SN. Figure \ref{fig:gal_cdf} shows the locations of three SNe in the CDF of NGC~1365 to illustrate this process.

We repeat this process for all SNe so that each SN now has an associated value measured from the CDF of its host galaxy. Then, following \citet{james06} and \citet{anderson12, anderson15}, we sort these CDF measurements for all SNe of a given type from lowest to highest. We use these sorted CDF values to build another cumulative distribution, now a distribution of associated host galaxy CDF values at the locations of SNe. We refer to these resulting distributions as ``SN distributions,'' which allow us to assess the degree to which, averaged across the whole sample, each SN type is more concentrated than, less concentrated than, or a close match to the light of their host galaxies.

\subsubsection{Example and Illustration} 
\label{sec:expectation}

\begin{figure*}[]
\centering
\includegraphics[width=.48\textwidth]{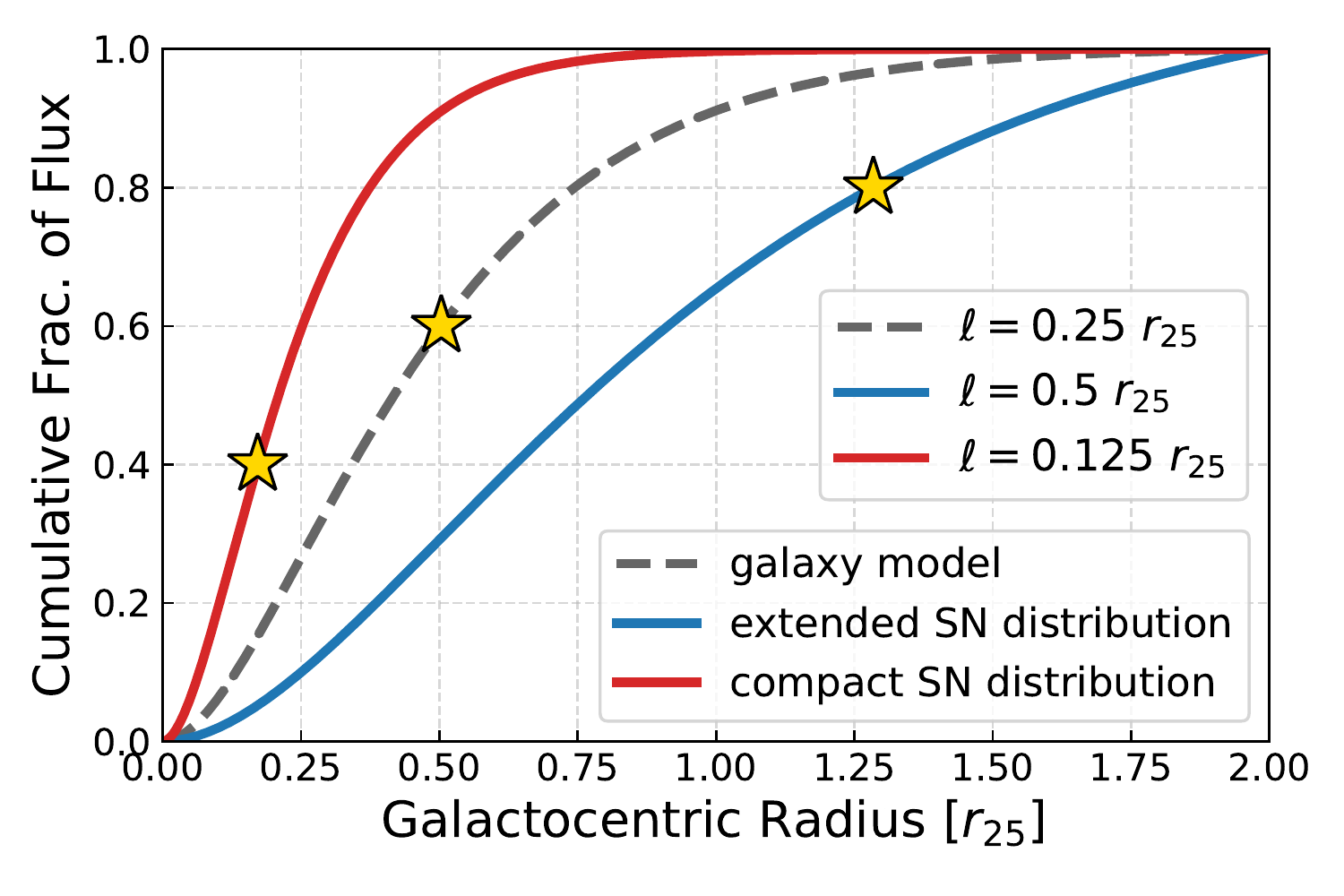}
\includegraphics[width=.48\textwidth]{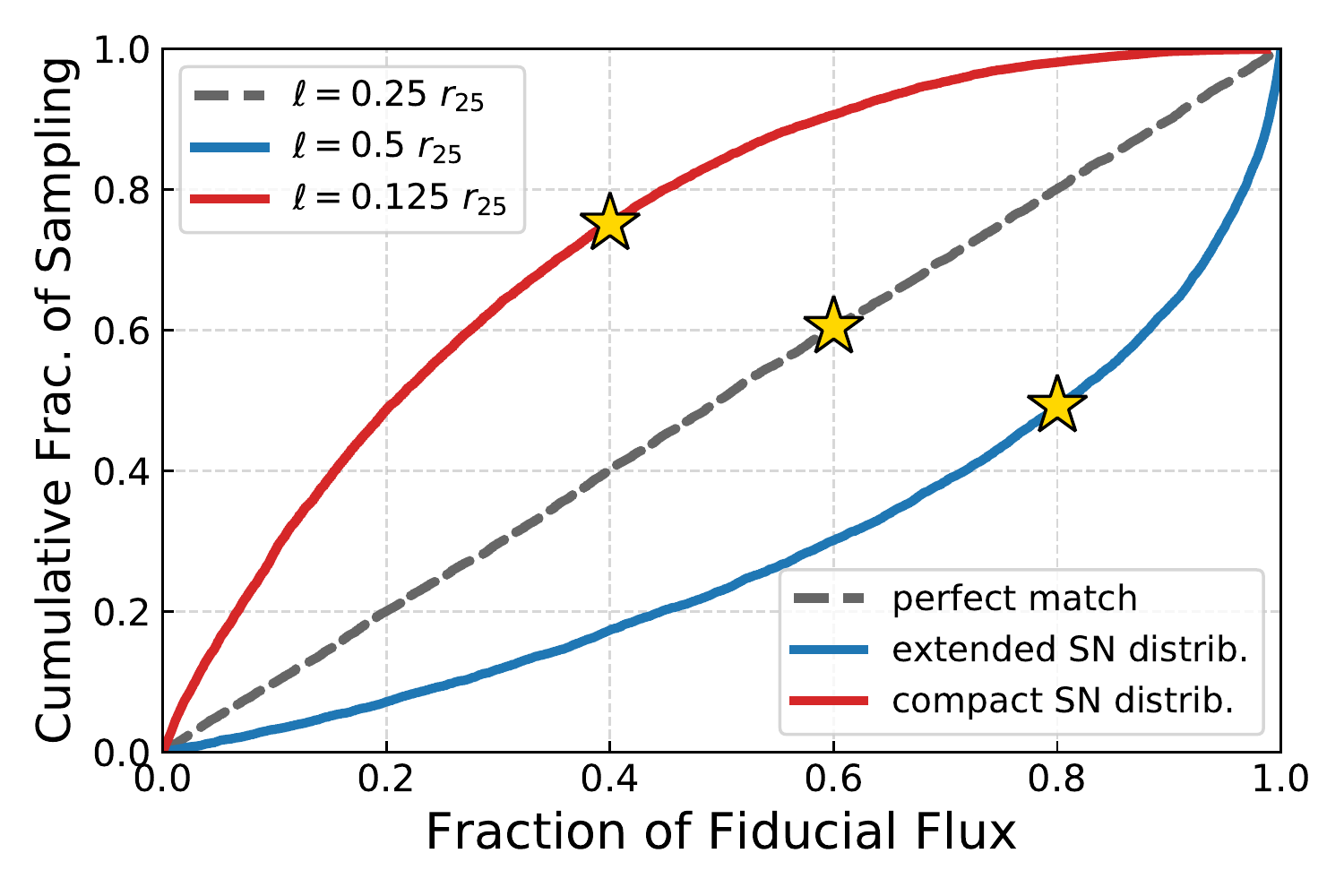}
\caption{\textbf{Illustration of our method for compact (red), extended (blue), and perfect match (black) SN distributions for a model galaxy.} The \textbf{left panel} shows the CDF for emission from a fiducial model galaxy that has scale length $\ell = 0.25~r_{25}$ in black (dashed line). We also plot model distributions that have half (red) and twice (blue) this fiducial scale length. If SNe are distributed like the light of this model galaxy, then their radial distribution will match the dashed black curve. If SNe are more concentrated than the light of their hosts, then their radial profiles will appear similar to the red curve, while the blue curve illustrates a more extended profile. The \textbf{right panel} shows the results of sampling each of these three cases with 1,000 SNe drawn from the appropriate probability distribution. The dashed black line shows a one-to-one perfect match between SNe distribution and the model (black in the left panel) galaxy profile. The case where SNe are twice as concentrated as the light from the galaxy curves up (red) relative to this perfect match case, while the more extended distribution of SNe curves down (blue). The three star symbols match between the two panels, and show the location of example draws for individual SNe.
\label{fig:illustrate}}
\end{figure*}

To illustrate this method, we model a fiducial galaxy that has an exponential radial intensity profile with a scale length of $\ell_{\rm fid} = 0.25~r_{25}$ (note that the exact choice of $\ell_{\rm fid}$ does not matter for this calculation). The dashed black curve in the left panel of Figure~\ref{fig:illustrate} shows the CDF of that galaxy. We then construct two other model distributions that could reflect more concentrated or more extended distributions of SNe compared to the galaxy model. The first has a radial profile that is more concentrated than the original model galaxy, with $\ell_{\rm con} = 0.5~\ell_{\rm fid} = 0.125~r_{25}$. The other has a more extended distribution, with $\ell_{\rm ext} = 2~\ell_{\rm fid} = 0.5~r_{25}$. The red curve in the left panel of Figure~\ref{fig:illustrate} shows the more concentrated CDF, while the blue curve shows the more extended CDF. 

In this illustration, we view the dashed black curve as the CDF of a particular band of light in a particular host galaxy. If SNe follow the distribution of this band, e.g., because they trace the progenitor's population well, then their radial profiles will match the profiles of their hosts. In this case, the locations of SNe will be drawn from the dashed black curve. However, if SNe are systematically more concentrated than the emission in this band, then their radial profiles will be above the dashed black curve, as illustrated by the red curve in this example. If SNe follow a systematically more extended distribution than the light in this band, then their radial profiles will be below the dashed black curve, a case illustrated by the blue curve.

The right panel of Figure~\ref{fig:illustrate} illustrates the SN distributions that would result from these ``compact'' (red), ``extended'' (blue), and ``perfect match'' (black) scenarios. To build the right panel, we randomly draw the galactocentric radius in the left panel with a probability set according to the corresponding compacted, extended, or matched CDF. These represent the galactocentric radii of SNe.

Next, we look up the value of the CDF of the corresponding galaxy model (i.e., the $y-$axis in the left panel) at the radius of the randomly selected ``SNe.'' We repeat this process 1,000 times, sort the recorded CDF values, and build the normalized CDF from those sorted CDF values. Therefore, the $x-$axis in the right panel is the CDF value from the $y-$axis in the left panel. The $y-$axis in the right panel is the cumulative number of random samplings, or SNe, normalized by the total number of these simulated SNe.

If the SNe follow the exact distribution of the host galaxy light, then this results in the dashed black diagonal one-to-one line in the right panel of Figure \ref{fig:illustrate}. However, if the distribution of SNe is more concentrated than the light of the host galaxies, as illustrated by the red curve, then the SN distribution is pulled up and to the left compared to the one-to-one line. If the distribution of SNe is more extended than the light of the host galaxies, as illustrated by the blue curve, then the SN distribution is pulled down and to the right compared to the one-to-one curve.

To illustrate the transformation from the left panel to the right panel of Figure~\ref{fig:illustrate}, we mark the locations of three simulated SNe as star symbols. These simulated SNe are located at galactocentric radii of $0.17~r_{25}$, $0.5~r_{25}$, and $1.28~r_{25}$ in simulated galaxies with CDFs modeled as the red, black, and blue curves, respectively. The CDF values at those radii are 0.4, 0.6, and 0.8, respectively. These CDF values mark the location of fiducial SNe in the $x-$axis of the right panel.

\subsubsection{Statistical Comparison to the ``Perfect Match'' Case}

\begin{deluxetable}{llc}
\tablecaption{\label{tab:statsummary} 
Summary of Statistical Tests for ``Perfect Match'' Between Band and SN Distribution}
\tablewidth{0pt}
\tablehead{
\colhead{Band} & \colhead{SN Type} & {Conclusion}
}

\startdata
&    SNe Ia (142)  & cannot reject\\
W1 (3.4~$\mu$m) &    SNe II (222) & reject \\
&    SNe Ib/c (108) & cannot reject\\
\hline
&    SNe Ia (142)  & cannot reject\\
W2 (4.5~$\mu$m) &    SNe II (222)   & inconclusive\\
&    SNe Ib/c (108) & cannot reject\\
\hline
&    SNe Ia (142)  & cannot reject\\
W3 (12~$\mu$m) &    SNe II (222)  & cannot reject\\
&    SNe Ib/c (108) & cannot reject\\
\hline
&    SNe Ia (42)  & cannot reject\\
W4 (22~$\mu$m) &    SNe II (77)  & cannot reject\\
&    SNe Ib/c (40) & cannot reject\\
\hline
&     SNe Ia (122)  & reject\\
NUV (231~nm) &     SNe II (194)  & inconclusive\\
&     SNe Ib/c (96) & reject\\
\hline
&     SNe Ia (83)  & reject\\
FUV (154~nm) &     SNe II (128)  & inconclusive\\
&     SNe Ib/c (66) & reject\\
\hline
&     SNe Ia (122) & cannot reject \\
$\Sigma_{\rm SFR}$(NUV$+$W3) &     SNe II (194) & cannot reject \\
&     SNe Ib/c (196) & cannot reject \\
\hline
&     SNe Ia (35) & cannot reject\\
$\Sigma_{\rm SFR}$(FUV$+$W4) &     SNe II (69)  & cannot reject\\
&     SNe Ib/c (36) & inconclusive \\
\enddata
\tablecomments{Inconclusive means at least one statistical test (KS or AD) results in the borderline between reject and cannot reject the null hypothesis (Appendix~\ref{sec:ks}).}
\end{deluxetable}

The ``perfect match'' line (black) in Figure \ref{fig:illustrate} corresponds to the case where the distribution of SN environments exactly matches the distribution of host galaxy light in that band. This would be expected if, e.g., the band is a perfect template for the progenitors of that SN type. As a result, this perfect match line represents a useful hypothesis to test our real measurements against.

In Appendix \ref{sec:ks}, we carry out a series of rigorous statistical comparisons for each band and each SN type. For each SN type in each WISE and GALEX band, we create 100 Monte Carlo realizations of this perfect match case and compare them to our real measurements using the Kolomogorov-Smirnov (KS) and Anderson-Darling (AD) tests. These tests allow us to ``reject'' or ``fail to reject'' the case of a perfect match between the SN radial distribution and the radial distribution of light from that band using a $p-$value threshold of $0.05$. We summarize the results in Table \ref{tab:statsummary}. A more detailed version of this table is in Appendix \ref{sec:ks} (Table \ref{tab:stats}).

These statistical tests indicate when the measurements appear unlikely to result from a true perfect match. They do not indicate how different the distributions are when they do not match. To gauge how much compact or extended the locations of SNe appear compared to the host galaxy flux, we will compare the SN distributions to the model cases in Figure~\ref{fig:illustrate}, while also using the spread of realizations in our Monte Carlo experiment as a \textit{de facto} gauge of uncertainty due to limited sample size (i.e., stochasticity).

\subsection{Results} 
\label{sec:results}

\begin{figure*}[]
\centering
\includegraphics[width=1\textwidth]{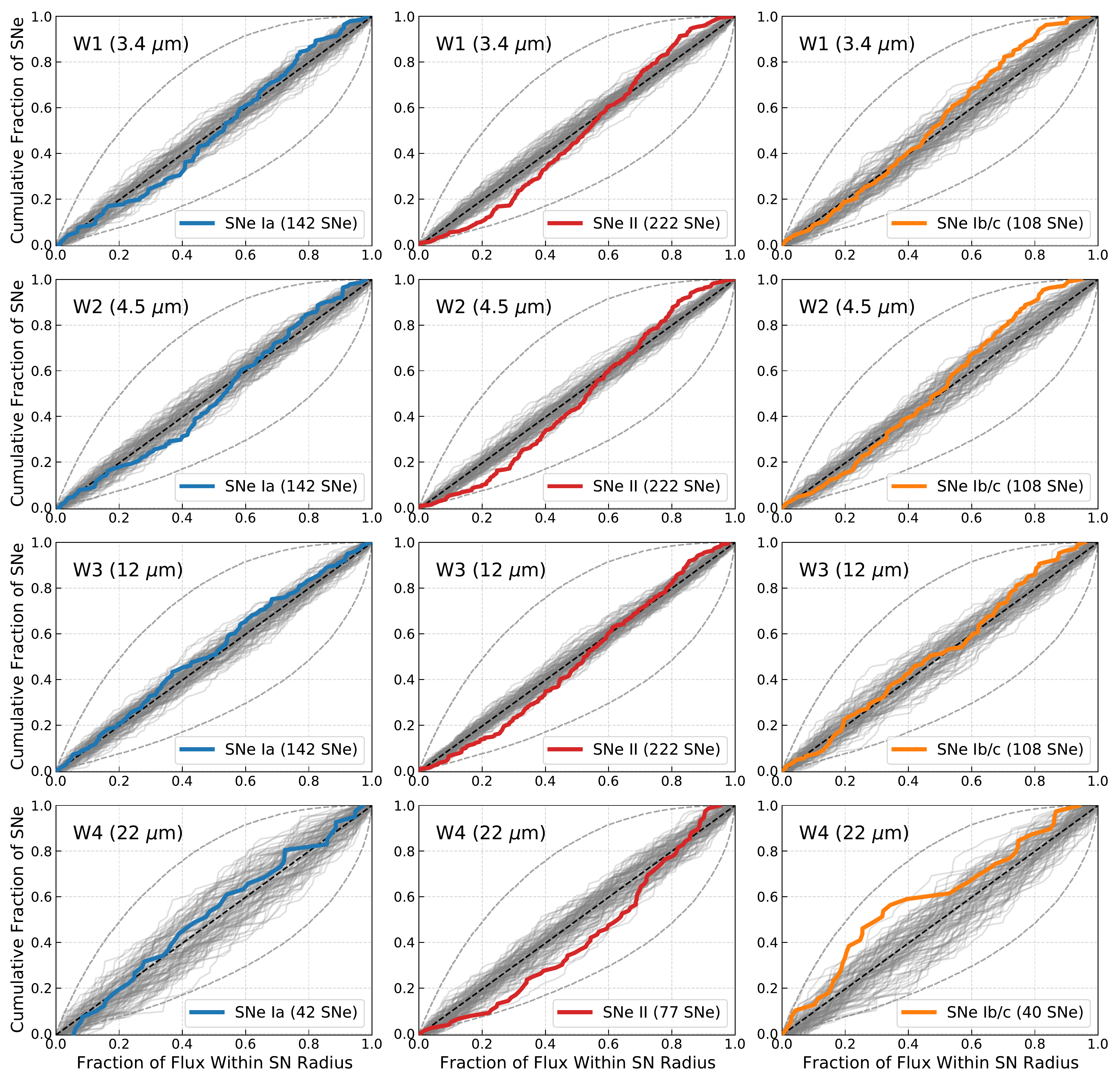}
\caption{{\bf Distributions that show how SN Ia, II, and Ib/c environments compare to the radial distribution of IR flux in their host galaxies.} Fraction of SNe Ia (blue), SNe II (red), and SNe Ib/c (orange) as a function of host galaxy IR (W1, W2, W3, and W4) emission at 2-kpc resolution enclosed within the SN's galactocentric radius. The dashed diagonal line shows the case where the SN environments are distributed exactly like the relevant band within each host galaxy. The solid gray curves are the resulting Monte Carlo simulations, which show the uncertainty in our measurements. A SN distribution that falls within the spread of the Monte Carlo simulations likely traces that band. The dashed gray curves are the model profiles from Figure~\ref{fig:illustrate}, where the curve pulled up and to the left represents a SN distribution that is $2\times$ more \textit{concentrated} than host galaxy emission, and the curve pulled down to the right represents a SN distribution that is $2\times$ more \textit{extended} than host galaxy emission.}
\label{fig:monte_carlo_1}
\end{figure*}

\begin{figure*}[]
\centering
\includegraphics[width=1\textwidth]{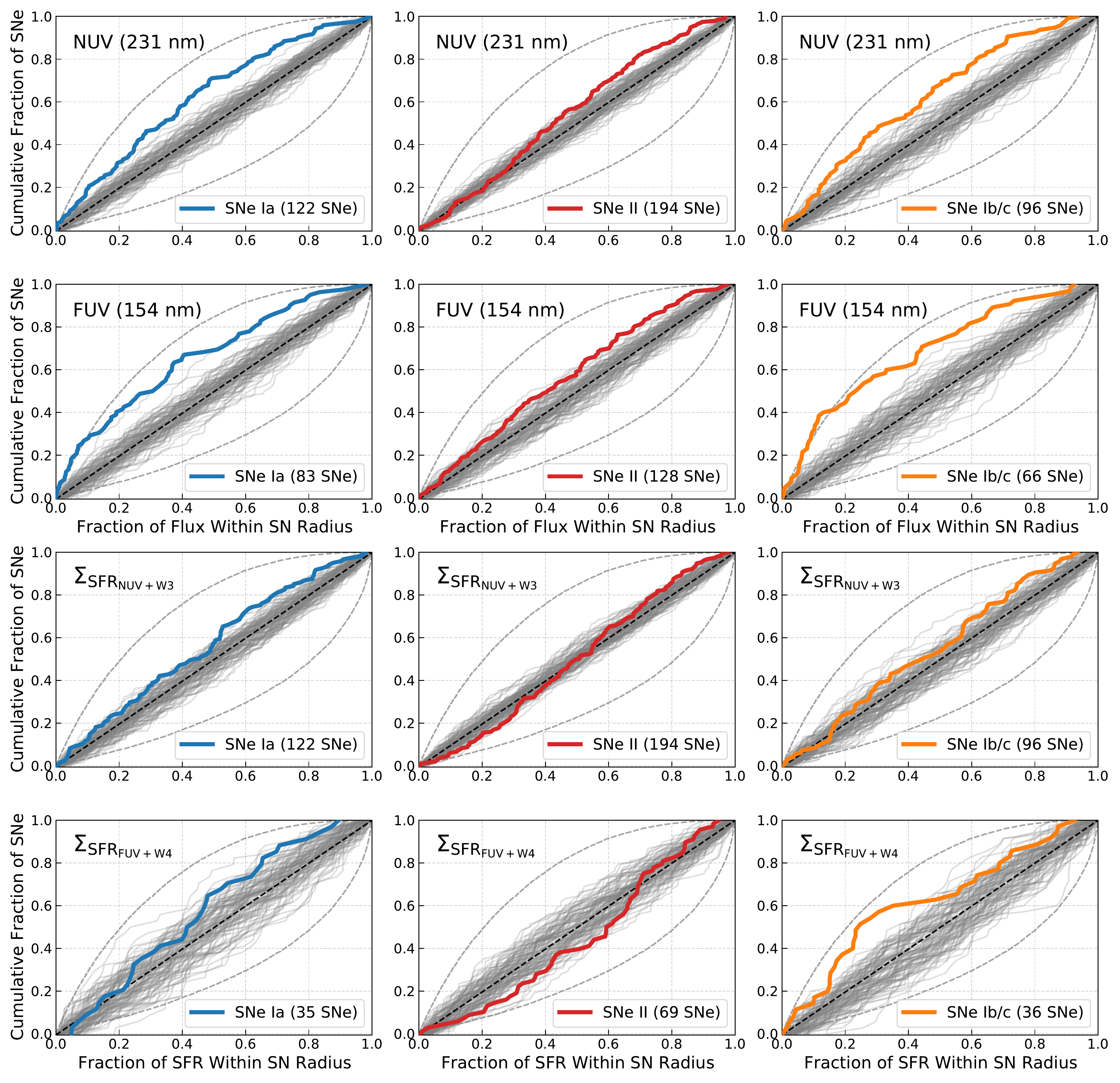}
\caption{{\bf Distributions that show how SN Ia, II, and Ib/c environments compare to the radial distribution of UV flux and $\Sigma_{\rm SFR}$ in their host galaxies.} Fraction of SNe Ia (blue), SNe II (red), and SNe Ib/c (orange) as a function of host galaxy UV emission and $\Sigma_{\rm SFR}$ at 2-kpc resolution enclosed within the SN's galactocentric radius. The dashed diagonal line shows the case where the SN environments are distributed exactly like the relevant band within each host galaxy. The solid gray curves are the resulting Monte Carlo simulations, which show the uncertainty in our measurements. A SN distribution that falls within the spread of the Monte Carlo simulations likely traces that band. The dashed gray curves are the model profiles from Figure~\ref{fig:illustrate}, where the curve pulled up and to the left represents a SN distribution that is $2\times$ more \textit{concentrated} than host galaxy emission, and the curve pulled down to the right represents a SN distribution that is $2\times$ more \textit{extended} than host galaxy emission.}
\label{fig:monte_carlo_2}
\end{figure*}

Figure \ref{fig:monte_carlo_1} shows the distributions of SNe Ia (blue), SNe II (red), and SNe Ib/c (orange) in our 2-kpc resolution maps of W1, W2, W3, and W4. We also plot the distributions of each SN type with respect to NUV, FUV, and $\Sigma_{\rm SFR}$ (as traced by FUV$+$W4 and NUV$+$W3) in Figure \ref{fig:monte_carlo_2}. The solid grey curves are the Monte Carlo samplings of a given band (including $\Sigma_{\rm SFR}$; as described in Appendix~\ref{sec:ks}).

In these plots, the $x{-}$axis indicates the fraction of host galaxy flux enclosed by the galactocentric radius of each SN site. The $y{-}$axis is the cumulative fraction of SNe that occur at a radius that encloses the amount of flux on the $x{-}$axis. For example, a point at $(x,y) = (0.4,0.75)$ indicates that 75\% of SNe occur within a galactocentric radius that encloses $40\%$ or less of the total galaxy flux.

As illustrated in Section~\ref{sec:expectation}, these figures show how well the distribution of each SN type follows the distribution of different bands across our galaxies. A perfect match would follow the dashed one-to-one line. When the measurement lies above the one-to-one line, the SNe are distributed, on average, more compactly than the distribution of that band of light. Conversely, a measurement below the one-to-one line indicates that the SNe appear more extended than that band of light.

For reference, we plot the compact and extended models from Figure~\ref{fig:illustrate} and Section~\ref{sec:expectation} as the dashed gray curves in Figures~\ref{fig:monte_carlo_1} and \ref{fig:monte_carlo_2}. As a reminder, these models illustrate cases where the SNe follow an exponential distribution with half (above curve) or twice (below curve) the scale length compared to the host galaxy light. Thus, comparing SN distributions with these curves aid in quantifying how much more concentrated or extended the SN distributions appear compared to the distribution of host galaxy emission.

Statistical analysis in Appendix~\ref{sec:ks} and by-eye analysis of the SN distributions in Figures~\ref{fig:monte_carlo_1} and \ref{fig:monte_carlo_2} lead us to the following observations regarding SN distributions relative to IR and UV emission.

\textbf{To first order, SNe follow the WISE surface brightness distributions of galaxies:} To first order, all types of SNe show distributions that resemble the near- and mid-IR light of their host galaxies. As Tables~\ref{tab:statsummary} and \ref{tab:stats} and Appendix~\ref{sec:ks} show, we cannot reject a perfect match between SNe Ia or Ib/c and any of W1, W2, W3, or W4. We also cannot reject a perfect match between SNe II and W3 or W4. While we can formally reject that SNe II are a perfect match to W1, Figure~\ref{fig:monte_carlo_1} shows that the curve still generally follows the one-to-one line and remains close to the Monte Carlo sampling. This is similar to SNe II in W2, which has an inconclusive result based on conflicting KS and AD test conclusions, which may mean SNe II borderline traces W2. Overall, our {\it first order} conclusion is that in our sample, the distributions of all three types of SNe generally follow the distribution of near- and mid-IR emission within galaxies.

This broad similarity might appear surprising. W1 and W2 are often used to trace mostly old stars, while W3 and W4 are sensitive to recent star formation, or hot dust emission. We expected SNe Ia to trace the overall stellar mass distribution and SNe II and Ib/c to track recent star formation, but not necessarily the reverse. However, as we saw in Section~\ref{sec:sfms} and Figure~\ref{fig:sfr_vs_star}, $\Sigma_{\rm SFR}$ and $\Sigma_\star$ show a strong correlation within our sample. Again, to first order, W1 and W2 are bright where W3 is also bright. Moreover, our calculations consider the CDFs for individual galaxies, and thus only distinguish between how well SNe trace a given band \textit{within} a galaxy. Our method is \textit{not} sensitive to, e.g., the overall ratio of W3 to W1 in a galaxy, which would trace the overall SFR/M$_\star$. Those calculations are more clearly seen in Section~\ref{sec:sfms}. As Figure \ref{fig:sfr_vs_star} shows, there can be a good correlation between $\Sigma_{\rm SFR}$ and $\Sigma_\star$ even in quiescent, early-type galaxies; the ratio of SFR to M$_\star$ is just lower overall.

With this caveat in mind, our results do agree well with previous works. The correspondence between SNe Ia and W1 and W2 agrees well with \citet{anderson15ia}. They found that the SN Ia distribution follows the $R-$band flux in galaxies, which also traces old stellar populations. The match between SNe II, Ib/c, and mid-IR-based tracers of recent star formation agrees with \citet{anderson09}, who found CC SNe to be radially distributed like H$\alpha$ emission, which traces SFR.

\textbf{Unclear results for SNe and W4 emission:} Beyond first order, the relation between the SN distributions and W4 (Figure~\ref{fig:monte_carlo_1}, and subsequently, Figure~\ref{fig:monte_carlo_2} for $\Sigma_{\rm SFR}$(FUV$+$W4)) appears less clear because of the smaller number of maps available for this band at fixed 2-kpc resolution. The effects of a smaller sample size are visualized in Figures~\ref{fig:monte_carlo_1} and \ref{fig:monte_carlo_2}, resulting in the wider spread among the Monte Carlo samplings (solid gray distributions). Details of how sample size affects these statistical results are further outlined in Appendix~\ref{sec:ks}.

\textbf{SNe show a compact distribution compared to UV emission:} To first order, all types of SNe appear to have a more compact distribution compared to both NUV and FUV emission, with all curves visibly above the one-to-one line in Figure \ref{fig:monte_carlo_2}. This reflects that both UV bands suffer from significant extinction in the inner parts of many galaxies. SNe still occur in these inner regions, but the UV emission observed by GALEX is attenuated by dust. This compact distribution for SNe Ia is in agreement with results from \citet{audcent-ross20} based on the galaxy sample from the Survey of Ultraviolet emission in Neutral Gas Galaxies (SUNGG) using the GALEX NUV band (O.I. Wong et al. in preparation). Furthermore, they found an agreement between SNe II and NUV, whereas our results are inconclusive or close to the borderline (Table~\ref{tab:statsummary}). This result is reflected by the KS test $p-$value being 0.08 for NUV and 0.12 for FUV, whereas the AD test $p{-}$value is 0.04 for NUV and 0.03 for FUV (Table~\ref{tab:stats}).

\textbf{SNe follow the $\Sigma_{\rm SFR}$ distribution of their host galaxies to varying degrees:} All SN types trace the $\Sigma_{\rm SFR}$(NUV$+$W3) of their host galaxies. More formally, we cannot reject a perfect match case between SNe Ia, II, Ib/c and $\Sigma_{\rm SFR}$(NUV$+$W3). Qualitatively, SNe II appear to trace the one-to-one line more closely than SNe Ia and Ib/c, which are a bit more concentrated than the $\Sigma_{\rm SFR}$(NUV$+$W3) of their hosts. Additionally, the SN Ia and Ib/c distributions appear to follow each other closely.

When comparing these results with the more robust tracer of the SFR, $\Sigma_{\rm SFR}$(FUV$+$W4), the statistical tests are affected by small sample size. To first order, all three SN types qualitatively appear to trace the $\Sigma_{\rm SFR}$(FUV$+$W4) of their galaxies, with SNe II being perhaps the closest to the perfect match case. Indeed, the outcomes of the statistical tests in Appendix~\ref{sec:ks} (see Table~\ref{tab:stats}) agree that we cannot reject the perfect match case between the one-to-one line and SNe Ia and II. 

Our results for SNe Ib/c against this direct tracer of the SFR, however, are inconclusive. The statistical tests do not all agree with each other and suggest that SNe Ib/c perhaps partially trace the $\Sigma_{\rm SFR}$ of their hosts. Visually, this inconclusive result is most likely due to the ``bump" in the SN Ib/c distribution at around $x \approx 0.3$, where the distribution becomes almost twice as compact as host galaxy emission (Figure~\ref{fig:monte_carlo_2}). While this bump is also visible in the SN Ib/c distributions in W4 and FUV, the results of the statistical tests are also influenced by the smaller sample size imposed by W4. Because of this discrepancy between the $\Sigma_{\rm SFR}$(FUV$+$W4) and $\Sigma_{\rm SFR}$(NUV$+$W3) results for SNe Ib/c, we deem SNe Ib/c to partially or marginally follow the $\Sigma_{\rm SFR}$ distribution of their galaxies.

These results are inconsistent with those presented in the \citet{anderson15} summary. For example, \citet{anderson12} found that SNe Ib/c track star formation (H$\alpha$) to a higher degree than SNe II and SNe Ia track H$\alpha$. Similarly, \citet{galbany18} also shows that SNe Ic occur in higher SFR, higher H$\alpha$ EW, and a higher fraction of the young ($0{-}40$~Myr) stellar population than other SN types. This supports the understanding that, of the CC SNe, SNe Ib/c have higher progenitor masses than SNe II. We note that in the case of both $\Sigma_{\rm SFR}$(FUV$+$W4) and $\Sigma_{\rm SFR}$(NUV$+$W3), our results are influenced by the attenuated UV emission at the center of galaxies. The results of each SN type against FUV and NUV bands exemplify this effect. We also compare our results for SNe Ib/c in $\Sigma_{\rm SFR}$(NUV$+$W3) with Figure~\ref{fig:sfr_vs_star}, where SNe Ib/c appear to be located in regions of high $\Sigma_{\rm SFR}$. These results may suggest that SNe Ib/c track something subtler than just the SFR.

\textbf{SNe Ib/c appear centrally concentrated:} Consistent with Section~\ref{sec:sfms}, we find that SNe Ib/c show the most distinct distributions compared to both emission from their hosts and the other SN types. The distributions of SNe Ib/c tend to appear above the one-to-one line, indicating that the SN Ib/c distributions are more concentrated than host galaxy emission in most bands. This concentration is not significant enough in W1, W2, and W3 to reject a perfect match between these bands and SNe Ib/c. Instead, this separation appears most prominent in the W4, NUV, FUV, and $\Sigma_{\rm SFR}$(FUV$+$W4) bands, where we can reject the perfect match case between SNe Ib/c and at least NUV and FUV. This prominent separation between SNe Ib/c and these four bands is consistent with the apparent concentration of SNe Ib/c to high $\Sigma_{\rm SFR}$ values in Figure~\ref{fig:sfr_vs_star}.

In general, previous works have found that CC SNe are more centrally concentrated than the light of their host galaxies, with SNe Ib/c dominating these central locations more so than SNe II, and SNe Ic dominating more than SNe Ib \citep{habergham10,habergham12,anderson15}. These previous studies further suggested that this centralization is more prominent in samples of interacting galaxies \citep{habergham10,habergham12}. 

Figures~\ref{fig:monte_carlo_1} and \ref{fig:monte_carlo_2} show a similar centralization of SNe Ib/c over SNe II and Ia, especially in the near-IR, W4, FUV, and $\Sigma_{\rm SFR}$(FUV$+$W4) bands. This concentration of SNe Ib/c may be the result of a metallicity effect, where SNe Ib/c are more concentrated in the higher-metallicity parts of galaxies, which are also more centrally located \citep{anderson09}. This argument is corroborated by \citet{galbany18}, who showed that SNe Ib and Ic are located in regions with higher metallicity than SNe II sites. Alternatively, \citet{habergham10,habergham12} suggested that IMF, binarity, and/or stellar rotation may also play important roles in the occurrence of SNe Ic.

It is worth noting that past studies have found significant differences between their SNe Ib and Ic populations \citep[e.g.,][]{anderson12}, with SNe Ic occupying more central regions of their hosts than SNe Ib in non-interacting galaxies \citep{habergham12}. We check for this difference by splitting our own SNe Ib and Ic populations but we found no statistical significant differences between them. Thus, in favor of a larger sample size, we continue to analyze the SN Ib/c population as a whole and simply note how they collectively appear more concentrated in their host galaxies than SNe Ia and II.

\textbf{Caveats and limitations of this approach:} This structural comparison has a few important caveats. First, we only measure the degree to which the SN distributions track the light in any given band \textit{within} a galaxy. This measurement is not sensitive to overall variations in, e.g., the rate of SNe per unit SFR \textit{among} galaxies. The analysis in Section~\ref{sec:sfms} explores that aspect.

Second, we only work with radial profiles. We adopt this approach because of the limited resolution of the data, the need to improve the signal-to-noise, particularly at W4, and the need to interpolate over masked regions, especially at W1 and W2. But this approach, along with the limited angular resolution of the IR and UV data, limits the sharpness of our measurements. We have emphasized our large sample size, but this work could be repeated for a smaller sample size at higher physical resolution to gain a sharper view of exactly where a smaller sample of SNe occur within galaxies.

Finally, our analysis matches the physical resolution of the multiwavelength data among all bands, but we do not apply any similar processing to the SN locations. In principle, this implies that even for a perfect match, we expect the SNe to appear slightly more compact relative to the light, but this should be a small effect.

\section{Summary} 
\label{sec:conclusions}

We characterize the local, 2-kpc environments of 472 SNe drawn from the Open Supernova Catalog \citep{guillochon17}. These SNe are located inside 359 galaxies that were observed by WISE, and often also GALEX, and processed into an atlas of near-IR ($3.4$ and $4.5\micron$), mid-IR ($12$ and $24\micron$) and UV maps by \citet{leroy19}.

From those UV and IR maps, we estimate $\Sigma_{\rm SFR}$ and $\Sigma_\star$ at the locations of SNe and place them in the $\Sigma_{\rm SFR}{-}\Sigma_\star$ space (Figure~\ref{fig:sfr_vs_star}). By separating into each SNe type and comparing $\Sigma_{\rm SFR}$ and $\Sigma_\star$ measurements at SN sites with respect to all regions inside the host galaxies, we place SN environments relative to the ``resolved star forming main sequence'' (RSFMS, i.e., the locus relating $\Sigma_{\rm SFR}$ and $\Sigma_\star$ for actively star forming regions of galaxies). To our knowledge, this is the first time such a measurement has been made. We find that:

\begin{enumerate}
    \item Core collapse (CC) SNe (consist of SNe II and Ib/c) largely follow the RSFMS and appear in regions of galaxies with high specific star formation rate, $\Sigma_{\rm SFR}/\Sigma_\star$ (Figure~\ref{fig:kde}). Furthermore, the environments of SNe Ib/c appear displaced to slightly higher $\Sigma_{\rm SFR}$ relative to SNe II.
    
    \item By contrast, SNe Ia appear at a wide range of $\Sigma_{\rm SFR}/\Sigma_\star$ and are often located well below the star forming main sequence in elliptical galaxies and the quiescent parts of late-type galaxies (Figure~\ref{fig:ia-cc-kde}). These trends largely agree with current hypotheses regarding the progenitors of different types of SNe and are in line with previous empirical work.
\end{enumerate}

We next study the radial distribution of SNe to characterize these 2-kpc environments in greater detail. Our SN sample spreads across many galaxies; therefore, we need an approach that controls for variations in the internal structure of these galaxies. We follow the method presented in \citet{anderson15}, measuring locations of SNe and then comparing these to the radial cumulative distribution function of near-to-mid IR light, UV light, and star formation rate in each host galaxy. We compare these distributions to show how well the locations of SNe Ia, II, and Ib/c follow the radial distributions of UV and IR light of their host galaxies (Figures~\ref{fig:monte_carlo_1} and~\ref{fig:monte_carlo_2}). This allows us to explore how each SN type is associated with the dust, young stars, $\Sigma_{\rm SFR}$, and stellar mass distribution of their hosts. We find that:

\begin{enumerate}

    \item To first order, SNe follow the radial distributions of both near-IR and mid-IR light. Although SNe~Ia are expected to follow the stellar mass distribution (as traced by near-IR light), the finding that SNe~Ia also follow mid-IR light (which is sensitive to recent star formation) is surprising. This may be due to the correlation between $\Sigma_{\rm SFR}$ and $\Sigma_\star$ within the host galaxies that we saw in Figure~\ref{fig:sfr_vs_star}. 
    \item To first order, CC SNe follow the mid-IR bands. This is expected because mid-IR light should show associations to tracers of recent star formation. Because mid-IR light also traces dust, the association between SNe Ib/c and these bands aligns with SN Ib/c progenitors needing stellar winds to strip their outer envelopes (and these stellar winds are more efficient in high-metallicity stars). Beyond first order, we find small statistical differences between SNe II and near-IR light.
    \item SNe show a more compact radial distribution compared to UV emission. This reflects that UV bands suffer from significant extinction in the inner parts of many galaxies.
    \item SNe Ia and II follow the radial distribution of $\Sigma_{\rm SFR}$ (UV+mid-IR) of their host galaxies to varying degrees. This result is expected for SNe~II, whose progenitors are young, massive stars that often die in or near the star-forming regions in which they were born. However, this result is not obvious for SNe~Ia. This may be related to the prompt-channel of SN~Ia progenitors \citep[e.g.,][]{MANNUCCI05}, which is correlated with star formation.
    \item SNe Ib/c marginally trace the radial distributions of $\Sigma_{\rm SFR}$ of their host galaxies. This association with $\Sigma_{\rm SFR}$ agrees with their progenitors being young, massive stars. However, the association between SNe Ib/c and $\Sigma_{\rm SFR}$ is not as clear-cut as the association between SNe II and $\Sigma_{\rm SFR}$. This may reflect the apparent concentration of SNe Ib/c toward the central, high-metallicity regions of their hosts. It is also possible that SNe Ib/c may trace something subtler than just the SFR. We also see this hinted in Figures~\ref{fig:sfr_vs_star} and \ref{fig:kde}, where SNe Ib/c are more concentrated toward higher values of $\Sigma_{\rm SFR}$.
    
\end{enumerate}

Future studies will be needed to further constrain the environmental properties of SNe. Using this sample from the OSC with H$\alpha$ will allow for direct comparison with results presented in \citet{anderson15}, providing insight between the two star formation tracers of H$\alpha$ and our constructed $\Sigma_{\rm SFR}$(UV$+$IR) band. Finer resolutions may be used to probe deeper into the immediate environments of SNe, though for now pose a risk of limiting galaxy sample sizes. Far-future surveys will be able to provide these finer resolutions while retaining a large sample of SNe, which will only continue to grow thanks to high-cadence, all-sky SN searches.

\acknowledgments

We thank the anonymous referee for constructive feedback that greatly improved this work. SAC, DU, and AKL acknowledge helpful discussions with the OSU Supernova Discussion Group, including Patrick Vallely, John Beacom, and Christopher Kochanek. SAC also acknowledges Jiayi Sun for Python and other technical help. This work was partially carried out through the OSU Department of Astronomy's Summer Undergraduate Research Program.

This work is supported by NASA ADAP grants NNX16AF48G and NNX17AF39G and National Science Foundation (NSF) grant No.~1615728. SAC is partially supported by the OSU College of Arts and Sciences Undergraduate Research Scholarship. The work of SAC, DU, and AKL is partially supported by NSF under grant Nos.~1615105, 1615109, and 1653300. TGW acknowledges funding from the European Research Council (ERC) under the European Union’s Horizon 2020 research and innovation programme (grant agreement No.~694343). EWK acknowledges support from the Smithsonian Institution as a Submillimeter Array (SMA) Fellow. LAL acknowledges support by a Cottrell Scholar Award from the Research Corporation of Science Advancement.

This publication makes use of data products from the Wide-field Infrared Survey Explorer, which is a joint project of the University of California, Los Angeles, and the Jet Propulsion Laboratory/California Institute of Technology, funded by the National Aeronautics and Space Administration. This work is based in part on observations made with the GALEX. GALEX is a NASA Small Explorer, whose mission was developed in cooperation with the Centre National d’Etudes Spatiales of France and the Korean Ministry of Science and Technology. GALEX is operated for NASA by the California Institute of Technology under NASA contract NAS5-98034.

We acknowledge the usage of the HyperLeda database (\url{http://leda.univ-lyon1.fr}).~This research has made use of the NASA/IPAC Extragalactic Database (NED), which is funded by the National Aeronautics and Space Administration and operated by the California Institute of Technology (\url{https://ned.ipac.caltech.edu}).

\software{\texttt{Jupyter} \citep{kluyver16}, \texttt{Astropy} \citep{astropy13, astropy18}, \texttt{Pandas} \citep{mckinney10}, \texttt{NumPy} \citep{van11, harris2020}, \texttt{SciPy} \citep{virtanen20}, \texttt{Seaborn} \citep{waskom17}, \texttt{Matplotlib} \citep{hunter07}}

\clearpage
\bibliography{main.bib}
\bibliographystyle{aasjournal}

\appendix

\section{Statistical Comparison Between Supernova Types in $\Sigma_{\rm SFR}{-}\Sigma_{\star}$ Space}\label{sec:t-test}

In Section~\ref{sec:sfms}, we place SNe in $\Sigma_{\rm SFR}{-}\Sigma_{\star}$ space and compare these locations to the ``star-forming main sequence'' (Figure~\ref{fig:sfr_vs_star}). We then compare the distributions of different SN  types in $\Sigma_{\rm SFR}{-}\Sigma_{\star}$ to one another by performing the Student's t-test (assuming unequal variances; \citealt{student}). The null hypothesis of the t-test states that the median of two distributions are the same. When the t-test returns a $p-$value $< 5\%$, then we {\it reject} the null hypothesis and conclude that the two distributions show statistically significant differences. We fail to reject the null hypothesis when the $p-$value $> 5\%$.

We perform the t-test between the distribution of each SN type as functions of $\log_{10}\Sigma_{\rm SFR}$ and $\log_{10}(\Sigma_{\rm SFR}/\Sigma_{\star})$ as seen in Figure~\ref{fig:kde}. We record the t-test results in Table~\ref{tab:t-test}. We reject the null hypothesis between all SN types except for the SN II and Ib/c distributions as a function of $\log_{10}(\Sigma_{\rm SFR}/\Sigma_{\star})$, where the $p-$value $= 7.8\%$. Therefore, we find that SNe II and Ib/c are distributed statistically the same as a function of $\log_{10}(\Sigma_{\rm SFR}/\Sigma_{\star})$. Furthermore, we also perform a t-test between SN Ia and CC SN distributions after grouping by their host galaxy morphological type (Figure~\ref{fig:ia-cc-kde}). The $p-$values of $\log_{10}\Sigma_{\rm SFR}$ between SNe Ia and CC SNe in early ($T<2$) and late-type ($T>2$) galaxies are 2.8\% and 2.2\%, respectively; the $p-$values in $\log_{10}(\Sigma_{\rm SFR}/\Sigma_{\star})$ between those SN types in early- and late-type galaxies are effectively zero. Thus, the difference between those populations is statistically significant.

\section{Statistical Comparison Between Supernova Distributions and Galaxy CDFs}
\label{sec:ks}

To see how well the radial distribution of SNe track the UV and IR light of their host galaxies, we statistically quantify their differences using the Kolmogorov-Smirnov \citep[{KS;}][]{massey51} and Anderson-Darling \citep[{AD;}][]{anderson11} two-sample tests. However, sample size plays a huge role in our measurements. We were concerned that this effect would not be readily captured by comparing our measurements to the diagonal one-to-one line (i.e., our ``perfect match''), because the diagonal one-to-one line is only expected for an infinite sample size. Therefore, we compared our measurements to a set of model distributions derived from a Monte Carlo (MC) simulation. We constructed these MC simulations assuming a perfect match and realized them with the same sample size as our SN data. Then, we applied the KS and AD tests to the MC and SN distributions. This allowed us to assess realistic $p-$values that accounted for our limited sample size.

We build the MC distributions of the host galaxies as follows. First, we randomly pick a number of pixels from our SN host galaxies. This number of pixels is equal to our SN sample size. In this random selection of pixels, we assign a probability to select each pixel by the flux in that pixel, so that in the limit of a large sample size we expect the outcome to show the same CDF as the underlying band being considered (i.e., one-to-one line).  In this way, we construct a case that assumes a ``perfect match'' but we model realistic stochasticity. For example, when building the W1 band CDF for SNe Ia, we randomly select 142 pixels to construct our sample. After selecting these pixels, we treat them as SN locations and build CDFs of the host galaxy emission using the same method used to analyze the SN distributions. 

We then repeat this process 100 times to realize 100 model distributions. The variation among these model distributions yields a realistic uncertainty when we compare it to the diagonal one-to-one line, which describes the assumption used to generate the model. If a SN distribution from Figures~\ref{fig:monte_carlo_1} and \ref{fig:monte_carlo_2} falls within the range of generated Monte Carlo distributions, then that SN distribution is not significantly different from the host galaxy emission distribution, at least at the $p \sim 0.01$ level, which is below our chosen significance threshold of $p < 0.05$.

These MC simulations appear as gray lines in Figures \ref{fig:monte_carlo_1} and \ref{fig:monte_carlo_2}, where they offer a way to distinguish the significance of measured deviations from a perfect match. We also use them to inform the results of our KS and AD tests.

\subsection{Kolmogorov-Smirnov Test}

The null hypothesis of the KS test states that the SNe and host galaxy emisison are drawn from the same distribution. The statistic value, $D$, measures the absolute maximum difference between two empirical cumulative distributions. Small $D$ means both distributions are similar to each other. The $p{-}$value is the probability that we {\it cannot} reject the null hypothesis. 

For each band and each SN type, we perform the KS test between the SN distribution and each Monte Carlo distribution (i.e.,~the gray lines in Figures \ref{fig:monte_carlo_1} and \ref{fig:monte_carlo_2}), resulting in 100 $D$ and $p-$values. We take the medians of those $D$ and $p-$values as our KS test results and record them in Table~\ref{tab:stats}. We use these results instead of testing against the exact diagonal line in Figures~\ref{fig:monte_carlo_1} and~\ref{fig:monte_carlo_2} because they are a more realistic representation of the host galaxy distribution, taking into account the limited sample size. We assign the uncertainties on the $D$ and $p-$values from the 16$^{\rm th}$ and 84$^{\rm th}$ percentiles of the $D$ and $p-$values of all 100 Monte Carlo distributions.

We can {\it reject} the null hypothesis that both distributions are the same if the median $p{-}$value is smaller than a given threshold, $\alpha$. For a one-tail test, the threshold for the $p{-}$value is typically $\alpha = 5\%$, which corresponds to two standard deviations away from the mean of a distribution (i.e., the 5\% tail of the distribution).

Equivalently, we can {\it reject} the null hypothesis at a level of $\alpha = 5\%$ if the median $D$ value satisfies
\begin{equation} \label{eq:D-value}
D > c(\alpha) \left(\frac{m + n}{m \times n}\right)^{0.5},
\end{equation}
where $c(\alpha) = c(0.05) = $ 1.385, and $m$ and $n$ are the sample sizes of the two distributions. We refer to the right hand side of Equation~\ref{eq:D-value} as the KS threshold value. The resulting $p-$values, $D$ values, thresholds, and conclusions of the KS test are recorded in Table \ref{tab:stats}. In general, both the $D$ and $p-$values agree to rejecting or not being able to reject the null hypothesis. 

\subsection{Anderson-Darling Test}

We support the KS test results by performing the Anderson-Darling (AD) $2$-sample test, which is more sensitive and gives more weight to the tails of distributions than the KS test. The AD and KS tests both share the same null hypothesis, which states that the two distributions are the same. The null hypothesis is rejected when the statistic value is larger than a critical value at a given significance level. In this case, the critical values are 0.325, 1.226, 1.961, 2.718, 3.752, 4.592, and 6.546, which correspond to significance levels of 25\%, 10\%, 5\%, 2.5\%, 1\%, 0.5\%, and 0.1\%.

We calculate the AD test using the same method as the KS test. We perform the AD test between the SN distributions and each Monte Carlo distribution. We then take the median of the statistic values and the significance levels at which the null hypothesis can be rejected. If the significance level is $<$ 5\%, then we reject the null hypothesis that the two distributions are the same. We record the results of the AD test in Table \ref{tab:stats}.

In general, both the KS and AD tests are in agreement with each other in rejecting or not rejecting the null hypothesis. There are four cases in which the results are inconclusive: SNe II in W2, NUV, and FUV, and SNe Ib/c in $\Sigma_{\rm SFR}$(FUV$+$W4). In the case of SNe II, the KS test returns $p-$value $> 5\%$ (cannot reject the null hypothesis), while the AD test gives $p-$value $< 5\%$ (can reject). For SNe Ib/c in $\Sigma_{\rm SFR}$(FUV$+$W4), both the KS and AD tests are borderline to the threshold, as the $p-$value of the KS test is 4\% and the significance level of the AD test is 5\%.

\subsection{Empirical Monte-Carlo Based Check}

To assess our confidence in our rejections and non-rejections, we calculate the fraction of $D$ values obtained from a KS test between the MC distributions and the exact one-to-one line ($D_{\rm MC}$) that are greater than the $D$ values between the SN distributions and the exact one-to-one line ($D_{\rm SN}$; see Table \ref{tab:stats}). If the fraction of $D_{\rm MC} > D_{\rm SN}$ is $\lesssim 5\%$, similar to the significance threshold for the other tests, then it is more unlikely the SNe follow the host galaxy distribution. This test is purely empirical, and so it provides a useful comparison to the $p-$value calculations from the KS test and the associated implicit assumptions (Equation~\ref{eq:D-value}). However, these test results could differ due to the small number of samples for some comparisons.

The $D_{\rm MC} > D_{\rm SN}$ analysis agrees with our conclusions from the other statistical tests in most cases.  However, our results differ for SNe Ib/c in W4, where the fraction of $D_{\rm MC} > D_{\rm SN}$ is 0\%. They also differ for SNe II in W3 and SNe Ia in $\Sigma_{\rm SFR}$(NUV$+$W3), with a fraction of 2\% and 3\%, respectively, where otherwise the KS and AD tests agree that we cannot reject the null hypothesis. The reason for this discrepancy is clear from Figures \ref{fig:monte_carlo_1} and \ref{fig:monte_carlo_2}, which shows that the measured SN distribution has a large deviation from the one-to-one line over a small range for all three cases we highlight here. These deviations differ from the Monte Carlo draws, which highlight the uncertainty centered around the one-to-one line but tend not to have these larger deviations. Since the KS test statistic measures only the largest deviation between two cumulative distributions, we measure a smaller fraction with $D_{\rm MC} > D_{\rm SN}$ for these cases relative to the $p$-values from the KS and AD tests. 

Overall, the general agreement between the KS, AD, and empirical tests demonstrates that our results reliably assess how well SN distributions offer a perfect match to the flux distribution.

\begin{splitdeluxetable*}{ccccccc|cccccc|ccBcccccccccccc}
\tabletypesize{\scriptsize}
\tablecaption{\label{tab:snsamp} The SN sample in 2-kpc resolution maps.}
\tablewidth{0pt}
\tablehead{
\multicolumn{7}{c|}{Supernova Data} & \multicolumn{6}{c|}{Pixel intensities at SN sites} & \multicolumn{2}{c}{$\log_{10}\Sigma_{\rm SFR}$ at SN sites} & \multicolumn{12}{c}{Host Galaxy Data} \\
\multicolumn{7}{c|}{} & \multicolumn{6}{c|}{[10$^{-1}$ MJy~sr$^{-1}$]} & \multicolumn{2}{c}{[$M_{\odot}~{\rm yr}^{-1}~{\rm kpc}^{-2}$] } & \multicolumn{12}{c}{} \\
\hline
\colhead{Name} & \colhead{Type} & \colhead{Disc.~Yr} & \colhead{RA$_{\rm SN}$ [$^\circ$]} & \colhead{Dec$_{\rm SN}$ [$^\circ$]} & \colhead{$r_{\rm SN}$ [$^\circ$]} & \multicolumn{1}{c|}{$z_{\rm SN}$} & \colhead{W1} & \colhead{W2} & \colhead{W3} & \colhead{W4}
& \colhead{NUV} & \multicolumn{1}{c|}{FUV} & \colhead{FUV$+$W4} & \colhead{NUV$+$W3} &  \colhead{PGC} & \colhead{Host Name} & \colhead{$T$} & \colhead{RA$_{\rm gal}$ [$^\circ$]} & \colhead{Dec$_{\rm gal}$ [$^\circ$]} & \colhead{$r_{\rm 25}$ [$^\circ$]} & \colhead{Incl. [$^\circ$]} & \colhead{PA [$^\circ$]} & \colhead{$z_{\rm gal}$} & \colhead{Dist.~[Mpc]} &\colhead{$\log_{\rm 10}(M_\star/M_{\odot})$} & \colhead{$\log_{\rm 10}({\rm SFR}/M_{\odot}~{\rm yr}^{-1})$}
}
\startdata
{SN2012dk}  & Ia & 2012 & 3.48  & $-$70.03 & 0.0082 & 0.015 & 0.61 & 0.34 
& 0.97     & \nodata & 0.17 & \nodata  & \nodata & $-$2.72 & PGC~926  & ESO50-G6 
& 6.0    & 3.5  & $-$70.02    & 0.013 & 43.06 & 88.64 & 0.014 & 56.96 & 9.83  & $-$0.18\\
{SN1994Z}   & II   & 1994 & 5.32  & $-$48.63 & 0.0086 & 0.012 & 0.18 & 0.08 
& $-$0.97  & \nodata & 0.09 & 0.06    & \nodata & $-$3.39 & PGC~1357 & NGC~87
& 9.9    & 5.31 & $-$48.63    & 0.007 & 38.08 & 5.47  & 0.012 & 49.3  & 9.21  & $-$0.23\\
{SN2004dd}  & II~P & 2004 & 6.97  & $-$1.81  & 0.0048 & 0.013 & 4.72 & 3.03 
& 16.7      & \nodata & 0.89 & \nodata  & \nodata & $-$1.83 & PGC~1715 & NGC~124
& 5.2    & 6.97 & $-$1.81     & 0.011 & 57.12 & 168.2 & 0.014 & 58.16 & 10.04 & 0.24\\
{SN2009hf}  & II   & 2009 & 9.34  & $-$19.95 & 0.018  & 0.013 & 1.75 & 1.02 
& 3.7      & \nodata & 0.21 & \nodata  & \nodata & $-$2.47 & PGC~2232 & NGC~175
& 2.2    & 9.34 & $-$19.93    & 0.017 & 54.36 & 101.9 & 0.013 & 54.6  & 10.78 & 0.26\\
{AT2018cdc} & II   & 2018 & 12.77 & $-$7.06  & 0.0172 & 0.006 & 0.23 & 0.1 
& 0.13     & 0.15   & 0.002 & 0.003    & $-$4.14  & $-$4.12  & PGC~2980 & NGC~274
& $-$2.8 & 12.76   & $-$7.06  & 0.011 & 37.81 & 157.6 & 0.006 & 19.5  & 9.88  & $-$0.52\\
{SN2014cx}  & II~P & 2014 & 14.95 & $-$7.57  & 0.0134 & 0.005 & 3.79 & 2.53 
& 12.1      & 25.1    & 1.57 & 1.13    & $-$1.76 & $-$1.72  & PGC~3572 & NGC~337
& 6.7    & 14.96   & $-$7.58  & 0.024 & 51    & 90    & 0.006 & 19.5  & 9.71  & 0.13\\ 
{SN2011dq}  & II   & 2011 & 14.95 & $-$7.57  & 0.0131 & 0.005 & 3.5  & 2.35 
& 10.8      & 23.2    & 1.53 & 1.1     & $-$1.77 & $-$1.74 & PGC~3572 & NGC~337
& 6.7    & 14.96   & $-$7.58  & 0.024 & 51    & 90    & 0.005 & 19.5  & 9.71  & 0.13\\  
{SN2013ct}  & Ia   & 2013 & 18.22 & 0.98   & 0.0052 & 0.004 & 4.98 & 2.85
& 3.5      & 5.5    & 0.6  & 0.4     & $-$2.28 & $-$2.17 & PGC~4367 & NGC~428 
& 8.6    & 18.23   & 0.982    & 0.023 & 47.89 & 114.3 & 0.004 & 14.8  & 9.4   & $-$0.32\\
{SN2010eb}  & Ia   & 2010 & 20.41 & 5.29   & 0.07   & 0.008 & 0.17 & 0.1  
& 0.15     & 1.48   & 0.001 & $-$0.0006 & $-$3.35 & $-$4.12 & PGC~4946 & NGC~488 
& 2.9    & 20.45   & 5.26     & 0.042 & 45.38 & 12.67 & 0.008 & 26.9  & 11.13 & 0.21 \\
{SN2016cyw} & II~P & 2016 & 21.48 & 16.6 & 0.0065 & 0.014 & 2.69 & 1.67
& 6.02     & \nodata & 0.42 & 0.36    & \nodata & $-$2.2 & PGC~5321 & IC~1702 
& 6.0    & 21.48   & 16.6     & 0.009 & 54.2  & 172.3 & 0.014 & 59.92 & 10.26 & $-$0.22\\
\enddata
\tablecomments{This table is available in its entirety in machine-readable form.
All SN data, including galaxy host name, are from the Open Supernova Catalog. Galaxy data come from $z$0MGS, HyperLEDA, and the NASA Extragalactic Database. Intensities are measured from the WISE and GALEX $z$0MGS images.}
\end{splitdeluxetable*}

\begin{deluxetable}{l c c c | c | l c c c | c}
\tablecaption{\label{tab:t-test} Student's (unequal variances) t-test results between each SN type in $\Sigma_{\rm SFR}{-}\Sigma_{\star}$ space.}
\tablehead{
\multicolumn{4}{c|}{$p{-}$value between $\log_{10}\Sigma_{\rm SFR}$ distributions} & \multicolumn{1}{c|}{Median} &
\multicolumn{4}{c|}{$p{-}$value between $\log_{10}(\Sigma_{\rm SFR}/\Sigma_{\star})$ distributions} &
\multicolumn{1}{c}{Median} \\
\colhead{} & \colhead{SNe~Ia} & \colhead{SNe~II} & \multicolumn{1}{c|}{SNe~Ib/c} & \multicolumn{1}{c|}{$\log_{10}\Sigma_{\rm SFR}$} & \colhead{} & \colhead{SNe~Ia} & \colhead{SNe~II} & \multicolumn{1}{c|}{SNe~Ib/c} & \multicolumn{1}{c}{$\log_{10}(\Sigma_{\rm SFR}$/$\Sigma_{\star})$}
}
\startdata
SNe Ia    & 1 & $4.63 \times 10^{-7}$ & $7.1 \times 10^{-9}$ & $-2.38$ & SNe Ia & 1 & $5.7 \times 10^{-19}$ & $2.9 \times 10^{-13}$ & $-10.45$ \\
SNe II    & $4.63 \times 10^{-7}$ & 1 & $3.2 \times 10^{-2}$ & $-2.12$ & SNe II & $5.7 \times 10^{-19}$ & 1 & $7.8 \times 10^{-2}$ & $-9.92$ \\
SNe Ib/c  & $7.1 \times 10^{-9}$ & $3.2 \times 10^{-2}$ & 1 & $-1.94$ & SNe Ib/c & $2.9 \times 10^{-13}$ & $7.8 \times 10^{-2}$ & 1 & $-9.98$ \\
\enddata
\tablecomments{We perform the Student's (unequal variances) t-test between SNe Ia, II, and Ib/c for values of $\log_{10}(\Sigma_{\rm SFR})$ and $\log_{10}(\Sigma_{\rm SFR}$/$\Sigma_{\star})$. We record the resulting $p-$values. All SN types have distinct $\log_{10}(\Sigma_{\rm SFR})$ distributions, and most have distinct $\log_{10}(\Sigma_{\rm SFR}$/$\Sigma_{\star})$ distributions. The exception is when we compare $\log_{10}(\Sigma_{\rm SFR}$/$\Sigma_{\star})$ distributions for SNe II and Ib/c. The $p-$value $= 7.8\%$, meaning we cannot reject the null hypothesis that the means of the $\log_{10}(\Sigma_{\rm SFR}$/$\Sigma_{\star})$ distributions for SNe II and Ib/c are the same. However, we note that (1) this is close to our threshold of $5\%$; and (2) when SNe IIb is included in the SNe II category rather than the SNe Ib/c category, the $p-$value between SNe II and Ib/c is $42\%$. This suggests that SNe IIb have lower $\log_{10}(\Sigma_{\rm SFR}$/$\Sigma_{\star})$ values than SNe Ib/c. We also record the median $\log_{10}(\Sigma_{\rm SFR})$ and $\log_{10}(\Sigma_{\rm SFR}$/$\Sigma_{\star})$ values for each SN type, showing that SNe II and Ib/c have very close median $\log_{10}(\Sigma_{\rm SFR}$/$\Sigma_{\star})$ values, whereas SNe Ib/c has a higher $\log_{10}(\Sigma_{\rm SFR})$ value than SNe II. SNe Ia are consistently concentrated to lower values of $\log_{10}(\Sigma_{\rm SFR}$/$\Sigma_{\star})$ and $\log_{10}(\Sigma_{\rm SFR})$.}
\end{deluxetable}

\begin{deluxetable*}{ll|ccc|cc|c|c}
\tablecaption{\label{tab:stats} KS test and Anderson-Darling test results between each SN distribution and the simulated host galaxy emission for the 2-kpc sample.}
\tablewidth{0pt}
\tablehead{
\colhead{} & \colhead{} & \multicolumn{3}{|c}{Kolmogorov-Smirnov Test} & \multicolumn{2}{|c|}{Anderson-Darling Test} & \colhead{KS Test (1:1 line)} & \multicolumn{1}{|c}{Conclusion} \\
\hline
\colhead{Band} & \colhead{SN Type} & \multicolumn{1}{|c}{$p-$value} & \colhead{$D-$value} & \colhead{threshold}  & \multicolumn{1}{|c}{statistic} & \colhead{significance} & \multicolumn{1}{|c}{$D_{\rm MC} > D_{\rm SN} $} &\multicolumn{1}{|c}{reject null} \\
\colhead{} & \colhead{(sample size)} & \multicolumn{1}{|c}{} & \colhead{} &
\colhead{value} & \multicolumn{1}{|c}{value} & \colhead{level} & \multicolumn{1}{|c}{(\%)} & \multicolumn{1}{|c}{hypothesis?}
}
\startdata
&    SNe Ia (142)  & $0.368^{+0.42}_{-0.21}$   & $0.109^{+0.02}_{-0.03}$  & 0.161 & $0.082^{+0.76}_{-0.6}$ & 0.25 & 19 &  cannot reject\\
W1 (3.4~$\mu$m) &   SNe II (222)  & $0.045^{+0.14}_{-0.04}$   & $0.131^{+0.03}_{-0.03}$  & 0.129 & $3.27^{+1.5}_{-1.3}$ & 0.015 & 0 &  reject \\
&    SNe Ib/c (108) & $0.249^{+0.29}_{-0.2}$  & $0.139^{+0.05}_{-0.03}$ & 0.185 & $1.08^{+1.8}_{-0.96}$ & 0.117 & 6 & cannot reject\\
\hline
&    SNe Ia (142)  & $0.295^{+0.31}_{-0.21}$    & $0.116^{+0.03}_{-0.03}$ & 0.161 & $0.267^{+0.67}_{-0.65}$ & 0.25 & 11 & cannot reject\\
W2 (4.5~$\mu$m) &    SNe II (222)   & $0.095^{+0.13}_{-0.08}$   & $0.117^{+0.03}_{-0.02}$ & 0.129 & $2.5^{+1.7}_{-0.9}$ & 0.031 & 1 & inconclusive\\
&    SNe Ib/c (108) & $0.249^{+0.38}_{-0.18}$   & $0.139^{+0.04}_{-0.04}$  & 0.185 & $1.2^{+1.5}_{-1.2}$ & 0.104 & 3 & cannot reject\\
\hline
&    SNe Ia (142)  & $0.593^{+0.29}_{-0.39}$  & $0.092^{+0.04}_{-0.02}$  & 0.161 & $-0.479^{+0.82}_{-0.39}$ & 0.25 & 35 & cannot reject\\
W3 (12~$\mu$m) &    SNe II (222)  & $0.226^{+0.23}_{-0.17}$  & $0.1^{+0.03}_{-0.02}$  & 0.129 & $1.1^{+1.2}_{-0.72}$ & 0.115 & 2 & cannot reject\\
&    SNe Ib/c (108) & $0.52^{+0.33}_{-0.33}$  & $0.111^{+0.04}_{-0.03}$ & 0.185 & $0.109^{+0.9}_{-0.64}$ & 0.25 & 29 & cannot reject\\
\hline
&    SNe Ia (42)  & $0.791^{+0.14}_{-0.36}$  & $0.143^{+0.05}_{-0.02}$  & 0.296 & $-0.511^{+0.61}_{-0.33}$ & 0.25 & 93 & cannot reject\\
W4 (22~$\mu$m) &    SNe II (77)  & $0.157^{+0.26}_{-0.11}$  & $0.182^{+0.04}_{-0.04}$  & 0.219 & $1.58^{+1.3}_{-1.2}$ & 0.072 & 2 & cannot reject\\
&    SNe Ib/c (40) & $0.165^{+0.24}_{-0.14}$   & $0.25^{+0.08}_{-0.05}$  & 0.304 & $1.05^{+1.7}_{-1.2}$ & 0.121 & 0 & cannot reject\\
\hline
&     SNe Ia (122)  & $0.005^{+0.02}_{-0.005}$ & $0.221^{+0.04}_{-0.03}$  & 0.174 & $6.79^{+3.1}_{-2.5}$ & 0.001 & 0 & reject\\
NUV (231~nm) &     SNe II (194)  & $0.08^{+0.17}_{-0.07}$   & $0.129^{+0.04}_{-0.03}$  & 0.138 & $2.31^{+2.2}_{-1.3}$ & 0.037 & 0 & inconclusive\\
&     SNe Ib/c (96) & $0.013^{+0.06}_{-0.01}$  & $0.229^{+0.05}_{-0.04}$ & 0.196 & $6.05^{+3.6}_{-2.7}$ & 0.001 & 0 & reject\\
\hline
&     SNe Ia (83)  & $0.003^{+0.01}_{-0.002}$ & $0.277^{+0.04}_{-0.04}$ & 0.211 & $7.8^{+2.3}_{-2.7}$ & 0.001 & 0 & reject\\
FUV (154~nm) &     SNe II (128)  & $0.119^{+0.22}_{-0.11}$   & $0.148^{+0.05}_{-0.03}$ & 0.17 & $2.67^{+2.9}_{-2}$ & 0.026 & 2 & inconclusive\\
&     SNe Ib/c (66) & $0.004^{+0.01}_{-0.004}$ & $0.303^{+0.06}_{-0.03}$ & 0.236 & $8.15^{+3.3}_{-2.6}$ & 0.001 & 0 & reject\\
\hline
&     SNe Ia (122)  & $0.164^{+0.24}_{-0.15}$    & $0.143^{+0.05}_{-0.03}$ & 0.174 & $1.79^{+2.3}_{-1.4}$ & 0.059 & 3 & cannot reject\\
$\Sigma_{\rm SFR}$ &     SNe II (194)  & $0.311^{+0.3}_{-0.18}$    & $0.098^{+0.02}_{-0.02}$ & 0.138 & $1^{+1.2}_{-0.89}$ & 0.126 & 20 & cannot reject\\
(NUV$+$W3) & SNe Ib/c (96) & $0.26^{+0.32}_{-0.21}$   & $0.146^{+0.05}_{-0.03}$ & 0.196 & $1.37^{+1.8}_{-1.4}$ & 0.089 & 17 & cannot reject\\
\hline
&     SNe Ia (35)  & $0.492^{+0.38}_{-0.38}$    & $0.2^{+0.09}_{-0.06}$ & 0.325 & $0.228^{+1.7}_{-0.72}$ & 0.25 & 19 & cannot reject\\
$\Sigma_{\rm SFR}$ &     SNe II (69)  & $0.249^{+0.5}_{-0.18}$    & $0.174^{+0.05}_{-0.06}$ & 0.231 & $0.675^{+1.3}_{-0.83}$ & 0.174 & 17 & cannot reject\\
(FUV$+$W4)&     SNe Ib/c (36) & $0.036^{+0.18}_{-0.03}$   & $0.333^{+0.06}_{-0.08}$ & 0.32 & $1.92^{+1.9}_{-1.3}$ & 0.052 & 1 & inconclusive\\
\enddata
\tablecomments{We perform the KS test between the SN distributions and each Monte Carlo simulation distribution (see Section \ref{sec:ks} and Figures \ref{fig:monte_carlo_1} and \ref{fig:monte_carlo_2}). 
The median of these 100 $p-$values and $D$ values are recorded in this table. To reject the null hypothesis, the $D$ value must be greater than a given threshold, $\alpha$. The $p{-}$value is the probability that the null hypothesis cannot be rejected. We support the KS test results with the Anderson-Darling test, which produces a statistic value and significance level at which the null hypothesis can be rejected. If the significance level is $<$ 5\%, then we reject the null hypothesis that the two distributions are the same. If the KS test and Anderson-Darling tests do not agree, then the results are deemed inconclusive. We also perform the KS test between each Monte Carlo distribution and the one-to-one line to get the value $D_{\rm MC}$, and another KS test between each SN distribution and the one-to-one line to get the value $D_{\rm SN}$. The column $D_{\rm MC} > D_{\rm SN}$ is the fraction of Monte Carlo distributions that have a $D$ value larger than the SN $D$ values, which assesses the confidence of our conclusions.}
\end{deluxetable*}

\end{document}